\documentclass[12pt,preprint]{aastex}
\usepackage{amssymb}

\shorttitle{Supermassive black hole binary mergers} 
\shortauthors{L.\'{A}. Gergely and P.L. Biermann} 

\begin{document}

\title{The spin-flip phenomenon \\
in supermassive black hole binary mergers}
\author{L\'{a}szl\'{o} \'{A}rp\'{a}d Gergely$^{1,2,3\star }$ and Peter L.
Biermann$^{4,5,6,7,8\ddag }$}

\affil{
$^{1}$Department of Theoretical Physics, University of Szeged, Tisza Lajos krt 84-86, 
Szeged 6720, Hungary\\
$^{2}$Department of Experimental Physics, University of Szeged, D\'{o}m t\'{e}r 9, 
Szeged 6720, Hungary\\
$^{3}$Department of Applied Science, London South Bank University, 103 Borough Road, 
London SE1 0AA, UK\\
$^{4}$Max Planck Institute for Radioastronomy, Bonn, Germany\\
$^{5}$Department of Physics and Astronomy, University of Bonn, Germany\\
$^{6}$Department of Physics and Astronomy, University of Alabama,
Tuscaloosa, AL, USA \\
$^{7}$Department of Physics, University of Alabama at Huntsville, AL, USA 
\\
$^{8}$FZ Karlsruhe and Physics Department, University of Karlsruhe, Germany\\
\small {$^\star$ E-mail: gergely@physx.u-szeged.hu\qquad $^\ddag$ E-mail:
plbiermann@mpifr-bonn.mpg.de} }

\begin{abstract}
Massive merging black holes will be the primary sources of powerful
gravitational waves at low frequency, and will permit to test general
relativity with candidate galaxies close to a binary black hole merger. In
this paper we identify the typical mass ratio of the two black holes but
then show that the distance where gravitational radiation becomes the
dominant dissipative effect (over dynamical friction) does not depend on the
mass ratio; however the dynamical evolution in the gravitational wave
emission regime does. For the typical range of mass ratios the final stage
of the merger is preceded by a rapid precession and a subsequent spin-flip
of the main black hole. This already occurs in the inspiral phase, therefore
can be described analytically by post-Newtonian techniques. We then identify
the radio galaxies with a superdisk as those in which the rapidly precessing
jet produces effectively a powerful wind, entraining the environmental gas
to produce the appearance of a thick disk. These specific galaxies are thus
candidates for a merger of two black holes to happen in the astronomically
near future.
\end{abstract}

\keywords{compact binaries, gravitational radiation, radio galaxies,jets}

\label{firstpage}

\section{Introduction}

The most energetic phenomenon that involves general relativity in the
observable universe is the merger of two supermassive black holes (SMBHs).
Therefore the study of these mergers may provide one of the most stringent
tests of general relativity even before the discovery and precise
measurement of the corresponding gravitational waves (see, e.g., Sch{\"{a}}%
fer 2005).

Most galaxies have a central massive black hole (Kor\-mendy \& Richstone
1995, Sanders \& Mirabel 1996, Faber et al. 1997), and after their initial
growth (for one possible example how this might happen, see Munyaneza \&
Biermann, 2005, 2006), their evolution is governed by mergers. Therefore,
the two central black holes also merge (see Zier \& Biermann 2001, 2002;
Biermann et al. 2000; Merritt \& Ekers 2002; Merritt 2003; Gopal-Krishna et
al. 2003, 2004, 2006; Gopal-Krishna \& Wiita 2000, 2006; Zier 2005, 2006,
2007). Before the two black holes get close, the galaxies begin to round
each other, distorting the shape of a radio galaxy fed by one or both of the
two black holes; thence the Z-shaped radio galaxies (Gopal-Krishna et al.
2003). When they merge, under specific circumstances to be clarified in this
paper, a spin-flip may occur. For a black hole nurturing activity around it,
the spin axis defines the axis of a relativistic jet, and therefore a
spin-flip results in a new jet direction: thence the X-shaped radio galaxies
(Rottmann 2001, Chirvasa 2001, Biermann et al. 2000, Merritt and Ekers
2002). In fact, observations suggest that all activity around a black hole
may result in a relativistic jet even for radio-weak quasar activity (Falcke
et al. 1996, Chini et al. 1989a, b). A famous color picture showing the past
spin-flip of the M87 black hole (Owen et al. 2000) clearly shows a weak
radio counter-jet, misaligned with the modern active jet by about 30$^{\circ
}$. The feature of the X-shaped radio galaxy jets is so common and yet very
short-lived that all radio galaxies may have been through this merger
(Rottmann 2001), and thus should have undergone a spin-flip. This can be
also deduced from the observation that many compact steep spectrum sources
show a misaligned double radio structure, where an inner pair of hot spots
is misaligned with an outer pair of hot spots (Marecki et al. 2003). We
conclude that theoretical arguments and observations consistently suggest
that black holes merge and result in a spin-flip.

From these and some other data we deduce a few basic tenets that the theory
needs to explain:

1. In the X-shaped radio galaxies the angles between two pairs of jets in
projection are typically less than 30 degrees. The real angles can be even
about 45$^{\circ }$. The jets are believed to signify the spin axis of the
more active (therefore presumably the more massive) black hole before the
merger and the spin axis of the merged black hole. Therefore, a substantial
spin-flip should have occurred.

2. In the X-shaped radio galaxies one pair of jets has a steep radio
spectrum. This implies that it has not recently been resupplied
energetically, it is an old pair of jets; and its synchrotron age is
typically a few $10^{7}$ years. The other pair of jets has a relatively flat
radio spectrum (this is the new jet; Rottmann 2001). Radio continuum
spectroscopy thus supports the spin-flip model.

3. Again, as Rottmann (2001) shows, the statistics of X-shaped radio
galaxies are such, that every radio galaxy may have passed through this
stage during its evolution. This matches with arguments based on
far-infrared observations that central activity in galaxies such as
starbursts and feeding of the activity of a central black hole, is often,
maybe always, preceded by a merger of galaxies (Sanders \& Mirabel 1996).

4. There is another critical observation of the spectrum of radio galaxies.
For many of them the radio spectrum has a low-frequency cutoff, suggesting a
cutoff in the energy distribution of the electrons at approximately the pion
mass (the electrons / positrons are decay products from pions, produced in
hadronic collisions; Falcke et al. 1995, Biermann et al. 1995, Falcke and
Biermann 1995a, 1995b, 1999, Gopal-Krishna et al. 2004). Hadronic collisions
with ensuing pion production at the foot ring of the radio jet occur
naturally and thermally in the case that the rotation parameter of the black
hole is larger than 0.95, and if the foot ring is an advection-dominated
accretion flow (ADAF) or radiatively inefficient accretion flow (RIAF; Donea
\& Biermann 1996, Mahadevan 1998, Gopal-Krishna et al. 2004). If this is
true for all radio galaxies, the spin of the black hole both before and
after the spin-flip must be more than $95~\%$ of the maximally allowed
value. This is a major constraint on the process of the spin-flip. If we
assume that this holds for all radio galaxies, then a fortiori it also holds
for those which have just undergone a binary black hole merger, and so their
spin ought to be high as well.

5. When two black holes merge, the emission of strong gravitational waves is
certain (Peters \& Mathews 1963, Peters 1964, Thorne 1979). Compact binaries
are driven by gravitational radiation through a post-Newtonian (PN) regime
(the inspiral), a plunge and a ring-down phase toward the final state. It is
commonly believed than the spin-flip phenomenon is likely to be caused by
the gravitational radiation escaping the merging system (Rottmann 2001,
Biermann et al. 2000, Merritt and Ekers 2002). Recent numerical work on the
final stages of the coalescence supports this (see Br{\"{u}}gmann 2007;
Campanelli et al. 2007a, b; Gonzalez et al. 2007a, b).

Therefore, it is mandatory to investigate what happens when the two black
holes get close to each other, and this we propose to treat in this paper.
We present here a model which allows \textit{to have a merger transition
going from a high-spin stage to another high-spin stage}, using mostly
physical insight from outside of the innermost stable orbit (ISO). In
contrast with available numerical simulation, our method, limited to a
certain typical range of mass ratios of the two black holes, has the
advantage that the evolution of the compact binary can be treated in the
framework of an analytical PN expansion with two small parameters.

In Section 2, we review the current state of observations on the masses of
supermassive galactic black holes, which roughly scale with the bulge masses
of their host galaxies. The observations suggest that the most massive black
holes have about $3\times 10^{9}$ solar masses ($M_{\odot }$) and the most
reliable determination of the low-mass central black hole (in our galaxy) is
about $3\,\times 10^{6}$ $M_{\odot }$ (Ghez et al. 2005, Sch\"{o}del \&
Eckart 2005). There is some evidence for central massive black holes of
slightly lower mass (Barth et al. 2005), but the error bars are very large.
This implies that the maximum mass ratio is about $10^{3}$. We carefully
analyze the statistics and argue in Section 2 that mass ratios in the range $%
3:1$ to $30:1$ cover most of the plausible range in mergers of galactic
central black holes. Roughly speaking, this means that typically one mass is
dominant by a factor of order 10. Therefore, we find that neither the much
discussed case of equal masses nor that of the extreme mass ratios (test
particles falling into a black hole) describes typical central galactic SMBH
mergers.

In Section 3, we study the relative magnitudes of the spin of the dominant
black hole and of the orbital angular momentum of the system. Their ratio
depends on two factors: the mass ratio and the separation of the binary
components (the inverse of which scales with the post-Newtonian parameter).
We show that for the typical mass ratio interval the orbital momentum left
when the system is reaching ISO is much smaller than the dominant spin. So
in the typical mass range case whatever happens during the plunge and
ring-down phases of the merger, in which the remaining orbital momentum is
dissipated, it cannot change essentially the direction of the spin. By
contrast, for equal mass mergers the orbital angular momentum dominates
until the end of the inspiral, while for extreme mass ratio mergers the
larger spin dominates from the beginning of the gravitational wave driven
merger phase.

In Section 4, we discuss the transition from the dynamical friction
dominated regime to the gravitational radiation dominated regime, in order
to establish the initial data for the PN treatment. The interaction of the
black holes with the already merged stellar environment generates a
dynamical friction when the separation of the black holes is between a few
parsecs (pc) and one hundredth of a pc. Gravitational radiation has a small
effect in this regime. Due to the dynamical friction, some of the orbital
angular momentum of the binary black hole system is transferred to the
stellar environment, such that the stellar population at the poles of the
system tends to be ejected and a torus is formed (Zier \& Biermann 2001,
2002, Zier 2006). This connects to the ubiquitous torus around active
galactic nuclei (AGNs), detected first in X-ray absorption (Lawrence \&
Elvis 1982, Mushotzky 1982), and later confirmed by optical polarization of
emission lines (Antonucci \& Miller 1985). Dynamical friction is enhanced as
in a merger the phase-space distribution is strongly disturbed by large
fluctuations of the mass distribution (Lynden-Bell 1967, Toomre \& Toomre
1972, Barnes \& Hernquist 1992, Barnes 2001). There had been a major worry
that the two black holes stall in their approach to each other (Valtonen
1996, Yu 2003, Merritt 2005, Milosavljevi{\'{c}} \& Merritt 2003a, b, Makino
\& Funato 2004, Berczik et al. 2005, 2006, Matsubashi et al. 2007) before
they get to the emission of gravitational waves; that the loss-cone
mechanism for feeding stars into orbits that intersect the binary black
holes is too slow. However, Zier (2006) has demonstrated that direct
interaction with the surrounding stars slightly further outside speeds up
the process, and so very likely no stalling occurs. Relaxation processes due
to cloud/star-star interactions are rather strong, as shown by Alexander
(2007), using observations of our galaxy. These interactions repopulate the
stellar orbits in the center of the galaxy. New work by Merritt et al.
(2007) is consistent with Zier (2006) and Alexander (2007). Also in a series
of papers Sesana, Haardt, and Madau have recently shown that even in the
absence of two-body relaxation or gas dynamical processes, unequal mass
and/or eccentric binaries with the mass larger than $10^{5}$ $M_{\odot }$
can shrink to the gravitational wave emission regime in less than a Hubble
time due to the binary orbital decay by three-body interactions in the
gravitationally bound stellar cusps (Sesana et al. 2006, 2007a, 2007b).
Finally, Hayasaki (2008) has considered the "last parsec problem" under the
assumption of the existence of three accretion disks: one around each black
hole and a third one, which is circumbinary. The circumbinary disk removes
orbital angular momentum from the binary via the binary-disk resonant
interaction, however, the mass transfer to each individual black hole adds
orbital angular momentum to the binary. The critical parameter of the mass
transfer rate is such that for SMBH binaries, it becomes larger than the
Eddington limit, thus these binaries will merge within a Hubble time by this
mechanism. The angular momentum transfer from orbit to disk was already
considered as a key physical concept in binary stars by Biermann \& Hall
(1973). All these recent works suggest that by one mechanism or another the
SMBHs will approach each other to distances smaller than approximately one
hundredth pc, when the gravitational radiation becomes the dominant
dissipative effect. In Section 4, we analyze the characteristic timescales
of the dynamical friction and gravitational radiation as function of the
total mass, stellar distribution radius and mass ratio of the compact binary
and we establish the values of the transition radius and PN parameter, for
which the gravitational radiation is overtaking dynamical friction.

In Section 5, we discuss the post-Newtonian evolution of the compact
binaries, following Apostolatos et al. (1994) and Kidder (1995). The new
element is the emphasis of the role of the mass ratio as a second small
parameter in the formalism. The leading order conservative effect
contributing to the change in the orientation of spins is the spin-orbit
(SO) coupling. The backreaction of the gravitational radiation, which is the
leading order dissipative effect below the transition radius, appears at one
PN order higher. We show here that for the characteristic range of mass
ratios the spin-flip occurs during the gravitational radiation dominated
inspiral regime, outside ISO. In the process we evaluate the timescales for
the change of the spin tilt as compared to the timescales of precessional
motion and gravitational radiation driven inspiral. As a by-product, we are
able to show that for the typical mass range the so-called transitional
precession occurs quite rarely.

We interpret and discuss the resulting model in Section 6. Here, we give a
tentative outline of the time sequence of the activity of two merging
galaxies, leading to an AGN episode of the primary black hole. A recent
review of the generic aspects of these galaxy nuclei as sources for ultra
high energy cosmic rays is in (Biermann et al. 2008).

Finally, we summarize our findings in the concluding remarks. Following our
arguments about the phase just barely before the merger we propose there
that the superwinds in radio galaxies (Gopal-Krishna et al. 2007,
Gopal-Krishna \& Wiita 2006) are in this stage, as the rapidly precessing
jet acts just like a powerful wind. The primary goal of our paper\ is to put
the derived physics into observational context, so as to allow tests to be
done in radio and other wavelengths.

\section{The relevant mass ratio range}

In Lauer et al. (2007), the mass distribution of galactic central black
holes is described, confirming earlier work, and also consistent with a
local analysis (Roman \& Biermann 2006). Arguments based on H\"{a}ring \&
Rix (2004), Gott \& Turner (1977), Hickson (1982), and Press \& Schechter
(1974) reasoning lead to a similar result, as does a recent observational
survey (Ferrarese at al. 2006b). Wilson \& Colbert (1995) also find a broken
power law. The probability for a specific mass ratio is an integral over the
black hole mass distribution, folded with the rate to actually merge
(proportional to the capture cross-section and the relative velocity for two
galaxies), e.g., isomorphic to the discussion in Silk \& Takahashi (1979)
for the merger of clumps of different masses. The black hole mass
distribution $\Phi _{BH}(M_{BH})$, the number of massive central black holes
in galaxies per unit volume, and black hole mass interval, can be described
as a broken power law, from about $m_{a}\simeq 3\times 10^{6}$ $M_{\odot }$
to about $m_{b}\simeq 3\times 10^{9}$ $M_{\odot }$, with a break near $%
m_{\star }\simeq 10^{8}$ $M_{\odot }$. The lower masses have been discussed
in some detail by Barth et al. (2005). The values of $m_{a},~m_{b}$ and $%
m_{\ast }$ imply that we have two mass ranges of a factor of $30$ each. The
masses above $10^{8}$ $M_{\odot }$ are rapidly becoming rare with higher
mass, so that the lower mass range is statistically more important. That
ratio range is then $1:1$ to $30:1$; while in the higher mass range the
maximal range of the masses is also $30:1$.

The mass of the central massive black hole scales with the mass of the
spheroidal component, as with the total mass of a galaxy (the dark matter),
see Benson et al. (2007). The rate of black hole mergers is some fraction of
all mergers of massive galaxies. If, as argued by Zier (2006) the approach
of the two black holes does not stall, then each merger of two massive
galaxies will inevitably lead to the merger of the two central black holes.
This is supported by the statistical arguments of Rottmann (2001), using
radio observations, that all strong central activity in galaxies may involve
a merger of two black holes. Therefore, observational evidence suggests that
black holes do merge, and do so on the rather short timescales of AGNs.

The interactions and mergers of galaxies clearly depend on the three angular
momenta: the two intrinsic spins, and the relative orbital angular momentum,
as well as on the initial distance and relative velocity of the two
galaxies. Once all these parameters are given, the evolution is quite
deterministic. The observations of Gilmore et al. (2007) strongly suggest,
that the initial seed galaxies are today's dwarf elliptical galaxies, all of
which are consistent with a lower bound to a common total mass of $5\,\times
10^{7}\;M_{\odot }$. This implies that all galaxies, and a fortiori all
central black holes, have undergone very many mergers.

The observations of Bouwens \& Illingworth (2006) and Iye et al. (2006)
strongly suggest that much of this merger history happened earlier than
redshift 6, perhaps mostly between redshifts $9$ and $6$. Each individual
merger runs along a well-defined evolutionary track, but all of these
mergers are completely uncorrelated with each other. Therefore, the ensemble
of very many mergers can be treated statistically, and this is what we
proceed to do, using the constant mass ratio between the spheroidal
component of galaxies and their central black holes. We thus use the merger
rate of galaxies as closely equivalent to the merger rate of the central
black holes.

The statistics of the mergers is given by the integral for the number of
mergers $N(q)$ per volume and time for a given mass ratio $q$, defined to be
larger than unity. This merger rate is the product of the distribution of
the first black hole with the distribution of the second black hole
multiplied by a rate $F$. The latter in principle depends on both the
cross-section and relative velocity of the two galaxies, the velocities
however are not very different, as the universe is not old enough for mass
segregation. The cross-section in turn depends on the two masses, thus $%
F=F(q,m)$. If we integrate for all cases, in which the first black hole is
less massive than the second black hole, we undercount by a factor of $2$,
and we have to correct for this factor. The general relationship is

\begin{equation}
N(q) = 2 \int_{m_{a}}^{m_{b} / q} \Phi_{BH} (m) \Phi_{BH} (q m) F(q, m) d m
\end{equation}

It is likely that the more massive black hole, and so the more massive host
galaxy, will dominate the merger rate $F$, so that it can be approximated as
a function of $qm$ alone, and a power law behavior with $F\sim q^{\xi }$
with $\xi >0$ should be adequate for a first approximation. To estimate $\xi 
$ roughly we just observe, that dwarf spheroidals have a core radius of a
few hundred pc (Gilmore et al. 2007), while our Galaxy has a core radius of
about $3$ kpc (Klypin et al. 2002), so a factor of $10$ in radius ($10^{2}$
in cross-section) for a factor of about $10^{4}$ in mass, thus the exponent
is likely to be approximately $1/2$; therefore a reasonable first estimate
for any cross-section is $\xi =1/2$. In this instance we use the approximate
equivalence of galaxy mergers with black hole mergers.

As the black hole mass distribution has a break at $q_{\ast }=30$, we use $%
\Phi _{BH}(m)\sim m^{-\tilde{\alpha}}$ for the first mass range, and $\Phi
_{BH}(m)\sim m^{-\tilde{\beta}}$ for the second. For the range $q$ from $1$
to $30$ we have as a dominant contribution

\begin{eqnarray}
N(q) &\sim &\int_{m_{a}}^{m_{\star }/q}\left( \frac{m}{m_{\star }}\right) ^{-%
\tilde{\alpha}}\left( \frac{mq}{m_{\star }}\right) ^{-\tilde{\alpha}}\left( 
\frac{mq}{m_{\star }}\right) ^{\xi }dm  \nonumber \\
&+&\int_{m_{\star }/q}^{m_{\star }}\left( \frac{m}{m_{\star }}\right) ^{-%
\tilde{\alpha}}\left( \frac{mq}{m_{\star }}\right) ^{-\tilde{\beta}}\left( 
\frac{mq}{m_{\star }}\right) ^{\xi }dm  \nonumber \\
&+&\int_{m_{\star }}^{m_{b}/q}\left( \frac{m}{m_{\star }}\right) ^{-\tilde{%
\beta}}\left( \frac{mq}{m_{\star }}\right) ^{-\tilde{\beta}}\left( \frac{mq}{%
m_{\star }}\right) ^{\xi }dm
\end{eqnarray}%
and for the case of $q$ above $30$ we have the contribution

\begin{equation}
N(q)\sim \int_{m_{a}}^{m_{b}/q}\left( \frac{m}{m_{\star }}\right) ^{-\tilde{%
\alpha}}\left( \frac{mq}{m_{\star }}\right) ^{-\tilde{\beta}}\left( \frac{mq%
}{m_{\star }}\right) ^{\xi }dm~.
\end{equation}

The various models shown in Lauer et al.(2007) show that a range of values
of $\tilde{\alpha}$ and $\tilde{\beta}$ is possible, with $\tilde{\alpha}$
ranging between approximately $1$ and $2$, and $\tilde{\beta}$ from $3$ to
larger values. Benson et al.(2007) propose $\tilde{\alpha}\approx 0.65$. We
adopt here the approximate values for $\tilde{\alpha}$ and $\tilde{\beta}$
of $1$ and $3$, to be cautious, and for $\xi $ we adopt $1/2$. With these
values the above integrands are monotonically decreasing functions and the
integrals are dominated by the lower limits. Thus, the four terms scale with 
$q$ as ${q}^{\xi -\tilde{\alpha}}$, ${q}^{-1+\tilde{\alpha}}$, ${q}^{\xi -%
\tilde{\beta}}$, and again ${q}^{\xi -\tilde{\beta}}$.

Let us consider the four terms: the first term is small galaxies merging
with small galaxies, and so not very interesting, as the cross-section is
low. However, for this distribution the number of mergers in the mass ratio
range $30:1$ to $3:1$ versus $3:1$ to $1:1$ is about $5$. The more extreme
mass ratios are more common. For the second term this ratio of mergers in
the two mass ratio ranges is about $14$. As this is massive galaxies merging
with smaller galaxies (above and below the break $m_{\star }$), this is the
most interesting case, and also quite common. The third term is almost
negligible, and the fourth term adds cases to the second term with more
extreme mass ratios, beyond $30:1$, and so emphasizes the large mass ratio
range.

So, among the relevant cases the rate of mergers of mass ratio of more than $%
3:1$ to those with a smaller mass ratio is in the range of $5:1$ to $14:1$,
about an order of magnitude. Focussing on those cases where one black hole
is at $10^{8}$ $M_{\odot }$ or larger, the ratio is larger than $14:1$.
Speculating that the exponent $\xi $ could be larger would enhance all these
effects; enlarging $\tilde{\alpha}$ would weaken them. Therefore we will
deal in the following with this much more common extended mass ratio range $%
30:1$ to $3:1$, which as will be shown, allows to use analytical methods.

\section{The spin and orbital angular momentum in the PN regime}

We assume the compact binary system to be composed of two masses $m_{i}$, $%
i=1,2$ , each having the spin $\mathbf{S}_{\mathbf{i}}$. By definition, the
characteristic radius $R_{i}$ of compact objects is of \ the same order of
magnitude that the gravitational radius $R_{G}=Gm_{i}/c^{2}$ (where $c$ is
the velocity of light and $G$ is the gravitational constant). Therefore, the
magnitude of the spin vector can be approximated as $S_{i}\approx
m_{i}R_{i}V_{i}\approx Gm_{i}^{2}V_{i}/c^{2}$, where $V_{i}$ is the
characteristic rotation velocity of the $i$th compact object \ As black
holes rotate fast due to accretion, $V_{i}/c$ is of order unity.
Equivalently we can introduce $S_{i}=(G/c)m_{i}^{2}\chi _{i}$, with $\chi
_{i}$ being the dimensionless spin parameter. Then maximal rotation implies $%
\chi _{i}=1$.

The PN expansion is done in terms of the small parameter 
\begin{equation}
\varepsilon \approx \frac{Gm}{c^{2}r}\approx \frac{v^{2}}{c^{2}},
\end{equation}%
where $m=m_{1}+m_{2}$ is the total mass and $v$ is the orbital velocity of
the reduced mass particle $\mu =m_{1}m_{2}/m$, which is in orbit about the
fixed mass $m$ (according to the one-centre problem in celestial mechanics).
The two expressions for $\varepsilon $ are of the same order of magnitude
due to the virial theorem. As in certain expressions odd powers of $v/c$ may
occur, it is common to have half-integer orders in the post-Newtonian
treatment of the inspiral of a compact binary system.

Whenever the masses of the two compact objects are comparable, either of $%
Gm_{i}/c^{2}r$ also represent one post-Newtonian order. However, as we have
argued before, for colliding galactic black holes it is typical that their
masses differ by 1 order of magnitude, so that we have a second small
parameter in the formalism. By choosing $m_{2}$ as the smaller mass, we can
also define the mass ratio 
\begin{equation}
\nu =\frac{m_{2}}{m_{1}}=\frac{1}{q}\in \left( 0,1\right) ~.
\end{equation}%
In the literature the symmetric mass ratio 
\begin{equation}
\eta =\frac{\mu }{m}\in \left( 0,\frac{1}{4}\right) ~
\end{equation}%
is also frequently employed. The two mass ratios are related as%
\begin{equation}
\eta =\frac{\nu }{\left( 1+\nu \right) ^{2}}~,
\end{equation}%
and for small $\nu $ we have $\eta =\allowbreak \nu -2\nu ^{2}+O\left( \nu
^{3}\right) $.

For the typical mass ratio range of SMBH binaries either $\eta $ or $\nu $
can be chosen as the second small parameter in the formalism. However, while
these stay constant, the PN parameter $\varepsilon $ evolves during the
inspiral toward higher values. Indeed, the separation of the components of
the binary with $m=10^{8}M_{\odot }$ evolves as 
\begin{equation}
r=\frac{Gm}{c^{2}\varepsilon }=\frac{r_{S}}{2\varepsilon }=\allowbreak
4.\,\allowbreak 781\,3\times 10^{-6}\frac{{pc}}{\varepsilon }{,}  \label{rep}
\end{equation}%
where $r_{S}$ represents the Schwarzschild radius. The interaction of the
galactic black holes with the stellar environment begins when the black
holes are a few kpc away from each other (then $\varepsilon \approx 10^{-8}$%
). The dynamical friction becomes subdominant at about $0.005$ pc (Zier \&
Biermann 2001, Zier 2006), when the gravitational radiation becomes the
leading dissipative effect. Thus, $\varepsilon =\varepsilon ^{\ast }\approx
10^{-3}$ is the value of the PN parameter for which the gravitational
radiation is driving the dissipation of energy and orbital angular momentum.
Then follows the inspiral stage of the evolution of compact binaries, which
continues until the domain of validity of the post-Newtonian approach is
reached, at few gravitational radii, at ISO. Further away a numerical
treatment is necessary in order to describe the plunge, which is finally
followed by the ring-down. The PN formalism can be considered valid until $%
\varepsilon \approx 10^{-1}$.

Theoretically, it is possible for a small $\nu $, that at certain stage of
the inspiral, the increasing $\varepsilon ^{1/2}$ becomes of the same order
of magnitude as $\nu $ and later on it even exceeds $\nu $. Such a situation
would shift the numerical value of several contributions to various physical
quantities into the range of higher or lower PN orders, depending of the
involved power of $\nu $. 
\begin{table*}[tbp]
\centering%
\begin{minipage}{140mm}
\caption{The evolution of the ratio 
$S_{1}/L\approx \protect\varepsilon^{1/2}\protect\nu ^{-1}$ 
in the range 
$\protect\varepsilon =10^{-3}\div 10^{-1}$ 
for various values of the mass ratio $\protect\nu $. }
\label{Table1}
\begin{center}
\begin{tabular}{lrlrl}
$S_{1}/L=\varepsilon^{1/2}\nu ^{-1}$ & 
					& 
$\varepsilon \approx 10^{-3}$ &  
					& 
$\varepsilon \approx 10^{-1}$ \\
\hline\hline
$\nu =1$ & $0.03$ & $\left( S_{1}\ll L\right) $ &
 \qquad $0.3$ & $\left( S_{1}< L\right) $ \\
$\nu =1/3$ & $0.1$ & $\left( S_{1}< L\right) $ & 
 \qquad $1$ & $\left(S_{1}\approx L\right) $ \\
$\nu =1/30$ & $1$ & $\left( S_{1}\approx L\right) $ & 
 \qquad $10$ & $\left(S_{1}> L\right) $ \\
$\nu =1/900$ & $30$ & $\left( S_{1}\gg L\right) $ & 
 \qquad $300$ & $\left(S_{1}\gg L\right) $
\end{tabular}
\end{center}
\end{minipage}
\end{table*}

The spin ratio (for similar rotation velocities $V_{1}\approx V_{2}$) can be
expressed as%
\begin{equation}
\frac{S_{2}}{S_{1}}\approx \left( \frac{m_{2}}{m_{1}}\right) ^{2}=\nu ^{2}{~.%
}  \label{S1S2}
\end{equation}%
The ratio of the spins to the orbital angular momentum becomes%
\begin{eqnarray}
\frac{S_{2}}{L} &\approx &\frac{Gm_{2}^{2}V_{2}/c^{2}}{\mu rv}=\left( \frac{%
Gm}{c^{2}r}\right) \left( \frac{c}{v}\right) \left( \frac{V_{2}}{c}\right) 
\frac{m_{2}}{m_{1}}\approx \varepsilon ^{1/2}\nu ~, \\
\frac{S_{1}}{L} &=&\frac{S_{2}}{L}\frac{S_{1}}{S_{2}}\approx \varepsilon
^{1/2}\nu ^{-1}~{.}  \label{S1L}
\end{eqnarray}%
We note that the approximations in the above formulae (\ref{S1S2})-(\ref{S1L}%
) are related only to the fact that we have assumed maximal rotation (thus $%
V_{i}/c\lesssim 1$). First, we note that the above ratios involving the
spins of the compact objects already contain $\varepsilon ^{1/2}$. Thus, the
counting of the inverse powers of $c^{2}$ is not equivalent with the PN
order, when compact objects are involved. Further, while the ratio $S_{2}/L$
is shifted toward higher orders by a small $\nu $ (therefore $S_{2}\ll L$
during all stages of the inspiral), the order of the ratio of the spin of
the dominant black hole to the magnitude of the orbital angular momentum is
not fixed. Indeed, it is determined by the relative magnitude of the small
parameters $\varepsilon $ and $\nu $. As $\varepsilon $ increases during the
inspiral, whenever $\nu $ falls in the range of $\varepsilon ^{1/2}$, the
initial epoch with $S_{1}<L$ is followed by $S_{1}\approx L$ and $S_{1}>L$
epochs (Table \ref{Table1}).

We have concluded in the previous section that the range of mass ratios $q$
between $3:1$ and $30:1$ is the most common. For such binaries\ the sequence
of the three epochs $S_{1}<L$, $S_{1}\approx L$, and $S_{1}>L$ is fairly
representative. We call this \textit{intermediate mass ratio mergers}, which
has to be contrasted with the case of equal mass mergers, where the orbital
angular momentum dominates throughout the inspiral; and with the case of
extreme mass ratio mergers (which we define as having mass ratios larger
than $30:1$), where the larger spin dominates from the beginning of the
inspiral to the end of the PN phase, as can be seen from our Table \ref%
{Table1}.

\section{The timescales}

The value $\varepsilon \approx 10^{-3}$ from which the PN analysis with the
gravitational radiation as the leading dissipative effect can be applied was
adopted in the previous section for a compact binary with total mass $%
m=10^{8}\,M_{\odot }$ and mass ratio $\nu =10^{-1}$. This was based on the
analysis in Zier \& Biermann (2001) and Zier (2006), where it was shown that
at around $5\times 10^{-3}$ pc gravitational radiation takes over from
dynamical friction in the interaction with stars in the angular momentum
loss of the black hole binary. Further arguments for the binary to reach the
gravitational wave emission regime were presented by Alexander (2007),
Sesana et al. (2006, 2007a, 2007b), and Hayasaki (2008).

In this section we raise the question, whether the value of the transition
radius (and the corresponding value of the PN parameter) depends on $m$ and $%
\nu $. In order to answer this, we compare the characteristic timescales of
the gravitational radiation and dynamical friction.

The timescale of gravitational radiation (as will be derived in Section 5)
is 
\begin{equation}
\frac{1}{t_{gw}}=-\frac{\dot{L}}{L}\approx \frac{32c^{3}}{5Gm}\varepsilon
^{4}\eta ~.
\end{equation}%
The timescale for the secondary black hole to lose angular momentum by
gravitational interaction with the surrounding stellar distribution is
(Binney \& Tremaine 1987) 
\begin{equation}
t_{fr}=\frac{v^{3}}{2\pi G^{2}m_{2}\rho _{distr}\Lambda }\left( \frac{\Delta
v}{v}\right) ^{2}~.
\end{equation}%
With the maximally allowable change in the velocity $\Delta v/v=1$ we find
the relevant \textit{full} timescale. Here, $\Lambda =\ln (b_{max}/b_{min})$
is the logarithm of the ratio of the maximal distance within the system,
divided by the typical distance between objects. The latter is large for
clouds, so the ratio is low and $\Lambda $ of the order unity, while for
stars or dark matter particles $\Lambda $ can be taken as $10\div 20$. For
the merger, the estimate based on clouds is more appropriate, thus,
following the reasoning of Binney \& Tremaine (1987) we adopt $\Lambda =3$.
The compact stellar distribution with density $\rho _{distr}$, radius $%
r_{distr}$, and mass $m_{distr}$ is of the same order in mass as the black
hole binary (Zier \& Biermann 2001, Ferrarese et al. 2006a) with a scale $%
r_{distr}$ of a few pc, and so under the assumption of a spherically
symmetric distribution we set%
\begin{equation}
\rho _{distr}=\frac{3m}{4\pi r_{distr}^{3}}~.
\end{equation}%
By employing the definitions of the PN parameter and mass ratio, we get%
\begin{equation}
\frac{1}{t_{fr}}=\frac{9G^{2}}{2c^{3}}\frac{m^{2}}{r_{distr}^{3}}\varepsilon
^{-3/2}\eta \left( 1+\nu \right) ~.
\end{equation}%
This gives the full timescale of the dynamical friction.

The two timescales become comparable for a PN parameter:%
\begin{equation}
\varepsilon ^{\ast }=K\left( \nu \right) \left( \frac{Gm}{c^{2}r_{distr}}%
\right) ^{6/11}~,
\end{equation}%
with $K\left( \nu \right) $ being a factor of order unity, defined as:%
\begin{equation}
K\left( \nu \right) =\left[ \frac{45}{64}\left( 1+\nu \right) \right]
^{2/11}\in \left( \allowbreak 0.938,~1.\,\allowbreak 064\right) ~.
\end{equation}%
corresponding to the distance $r^{\ast }$%
\begin{equation}
r^{\ast }=K^{-1}\left( \nu \right) \left( \frac{Gm}{c^{2}}\right)
^{5/11}r_{distr}^{6/11}~.
\end{equation}%
Notably, the dependence on the mass ratio is rather weak and in practice it
can be neglected. Inserting then for $r_{distr}=5$ pc and using as a
reference value for the mass $m=10^{8}~M_{\odot }$, we obtain $\varepsilon
^{\ast }\approx 10^{-3}$ and $r^{\ast }\approx 0.005$ pc in agreement with
the discussion in Zier \& Biermann (2001). We also note that in fact any
other reasonable value for $\Lambda $ and $\Delta v/v$ will give a factor of
order unity in Eq. (14), as this number arises by taking the power $2/11$.
The weak dependence on $\nu $ through $K$ is due to the same reason.

The value of the PN parameter thus scales with $m^{6/11}$ and radius $%
r_{distr}^{-6/11}$ as%
\begin{equation}
\varepsilon ^{\ast }\approx 10^{-3}\left( \frac{m}{10^{8}~M_{\odot }}\right)
^{6/11}\left( \frac{5{\ pc}}{r_{distr}}\right) ^{6/11}~,
\end{equation}%
while the transition radius scales with $m^{5/11}$ and $r_{distr}^{6/11}$ as 
\begin{equation}
r^{\ast }\approx 0.005~pc~\left( \frac{m}{10^{8}M_{\odot }}\right)
^{5/11}\left( \frac{r_{distr}}{5\ {\ }pc}\right) ^{6/11}~.
\end{equation}%
For $m=10^{9}M_{\odot }$ and for the same radius of stellar distribution
then $r^{\ast }\approx $ $0.01$ pc.

We conclude that both the dependence of the transition radius and the
corresponding PN parameter on the total mass and on the stellar distribution
radius are weak, while there is practically no dependence on the mass ratio.

\section{The inspiral of spinning compact binaries in the gravitational
radiation dominated regime}

In the first subsection of this section we present the conservative dynamics
of an isolated compact binary in a post-Newtonian treatment, emphasizing the
role of the second small parameter, as a new element. Then in the second
subsection we take into account the effect of gravitational radiation,
deriving how the spin-flip occurs for the typical mass ratio range. The
limits of validity of our results obtained by using these two small
parameters will be considered below in subsection 5.3.

\subsection{Conservative dynamics below the transition radius}

The interchange in the dominance of either $L$ or $S_{1}$ has a drastic
consequence on the dynamics of the compact binary. To see this, let us
summarize first the conservative dynamics, valid up to the second
post-Newtonian order. The constants of the motion are the total energy $E$
and the total angular momentum vector $\mathbf{J=L+S}_{\mathbf{1}}+\mathbf{S}%
_{\mathbf{2}}$ (Kidder et al. 1993). The angular momenta $\mathbf{L,~S}_{%
\mathbf{i}}$ are not conserved separately. The spins obey a precessional
motion (Barker \& O'Connell 1975, Barker \& O'Connell 1979):%
\begin{equation}
\mathbf{\dot{S}_{i}}=\mathbf{\Omega }_{\mathbf{i}}\times \mathbf{S_{i}}\ ,
\label{spinprec}
\end{equation}%
with the angular velocities given as a sum of the spin-orbit, spin-spin, and
quadrupole-monopole contributions. The latter come from regarding one of the
binary components as a mass monopole moving in the quadrupolar field of the
other component.

The leading order contribution due to the SO interaction (discussed in
Kidder et al. 1993, Apostolatos et al. 1994, Kidder 1995, Ryan 1996, Rieth
\& Sch\"{a}fer 1997, Gergely at al. 1998a, 1998b, 1998c, O'Connell 2004),
cause the spin axes to tumble and precess. The spin-spin (Kidder 1995,
Apostolatos 1995, Apostolatos 1996, Gergely 2000a, 2000b), mass quadrupolar
(Poisson 1998, Gergely \& Keresztes 2003, Flanagan \& Hinderer 2007, Racine
2008), magnetic dipolar (Ioka \& Taniguchi 2000, Vas\'{u}th et al. 2003),
self-spin (Mik\'{o}czi et al. 2005) and higher order spin-orbit effects
(Faye et al. 2006, Blanchet et al. 2006) slightly modulate this process.

The SO precession occurs with the angular velocities 
\begin{eqnarray}
\mathbf{\Omega }_{\mathbf{1}} &=&{\frac{G\left( 4+3\nu \right) }{2c^{2}r^{3}}%
}\mathbf{L_{N}}~,  \label{Om1} \\
\mathbf{\Omega }_{\mathbf{2}} &=&{\frac{G\left( 4+3\nu ^{-1}\right) }{%
2c^{2}r^{3}}}\mathbf{L_{N}}~,  \label{Om2}
\end{eqnarray}%
where $\mathbf{L_{N}}=\mu \mathbf{r}\times \mathbf{v}$ is the Newtonian part
of the orbital angular momentum. The total orbital angular momentum $\mathbf{%
L}$ also contains a contribution $\mathbf{L}_{\mathbf{SO}}$ (Kidder 1995),
which for compact binaries is of the order of $\varepsilon ^{3/2}L_{N}$ .

Due to the conservation of $\mathbf{J}$, the orbital angular momentum
evolves as

\begin{equation}
\mathbf{\dot{L}=}{\frac{G}{2c^{2}r^{3}}}\left[ \left( 4+3\nu \right) \mathbf{%
S_{1}}+{\left( 4+3\nu ^{-1}\right) }\mathbf{S_{2}}\right] \mathbf{\times L~.}
\end{equation}%
(By adding a correction term of order $\varepsilon ^{3/2}$ relative to the
leading order terms,$\ $we have changed $\mathbf{L_{N}}$ into $\mathbf{L}$
on the right-hand side of the above equation.) \ 

To leading order in $\nu $ we obtain:%
\begin{eqnarray}
\mathbf{\dot{S}_{1}} &=&{\frac{2G}{c^{2}r^{3}}}\mathbf{L}\times \mathbf{S_{1}%
}\ ,  \label{Sdot} \\
\mathbf{\dot{L}} &\mathbf{=}&{\frac{2G}{c^{2}r^{3}}}\mathbf{S_{1}\times L~.}
\label{Ldotconserv}
\end{eqnarray}%
(Again, an $\mathbf{L}_{\mathbf{SO}}$ term was added to $\mathbf{L_{N}}$ on
the right-hand side of Eq. (\ref{Sdot}), in order to have a pure precession
of $\mathbf{S_{1}}$.)

Thus, the leading order conservative dynamics gives the following picture:
the dominant spin $\mathbf{S_{1}}$ undergoes a pure precession about $%
\mathbf{L}$, while$\ \mathbf{L}$ does the same about $\mathbf{S}_{\mathbf{1}%
} $. Despite the precession (\ref{Om2}), the spin $\mathbf{S}_{\mathbf{2}}$
can be ignored to leading order, as its magnitude is $\nu ^{2}$ times
smaller than $S_{1}$, e.g., Eq. (\ref{S1S2}). By adding the vanishing terms $%
\left( {2G/}c^{2}r^{3}\right) \mathbf{S}_{\mathbf{1}}\times \mathbf{S_{1}}$
and $\left( {2G/}c^{2}r^{3}\right) \mathbf{L}\times \mathbf{L}$ to the
right-hand sides of Eqs. (\ref{Sdot}) and (\ref{Ldotconserv}), respectively,
we obtain%
\begin{eqnarray}
\mathbf{\dot{S}_{1}} &=&{\frac{2G}{c^{2}r^{3}}}\mathbf{J}\times \mathbf{S_{1}%
}\ ,  \label{SdotJ} \\
\mathbf{\dot{L}} &\mathbf{=}&{\frac{2G}{c^{2}r^{3}}}\mathbf{J\times L~.}
\label{LdotJcons}
\end{eqnarray}%
Thus, the precessions can also be imagined to happen about $\mathbf{J}$,
which represents an invariant direction in the conservative dynamics up to
2PN.

Higher order contributions to the conservative dynamics slightly modulate
this precessional motion. In fact, for both the spin-spin and
quadrupole-monopole perturbations an angular average $\bar{L}$ can be
introduced, which is conserved up to the 2PN order (Gergely 2000a). As $\bar{%
L}$ differs from $L$ just by terms of order 2PN, and $\bar{L}$ is conserved,
the real evolution of $\mathbf{L}$ differs from a pure precession only
slightly.

Finally, we note that as the SO precessions are 1.5 PN effects and the
gravitational radiation appears at 2.5 PN, at the transition radius the SO
precession timescale is $\varepsilon ^{-1}$ times shorter than the timescale
of dynamical friction. The modifications induced by the precessions in the
angular momentum transfer toward the stellar environment will be discussed
elsewhere (Zier et al. 2009, in preparation).

\subsection{Dissipative dynamics below the transition radius}

\begin{figure*}[tbp]
\centering\includegraphics[height=8cm]{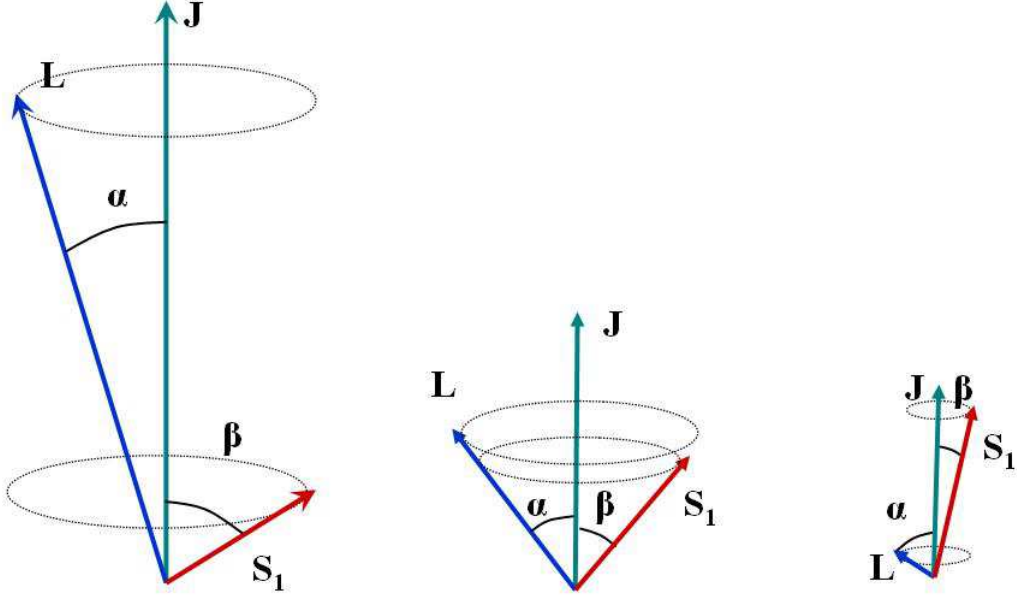}
\caption{The old jet points in the direction of the original spin $\mathbf{%
S_{1}}$. When the two black holes approach each other due to the motion of
their host galaxies, a slow precessional motion of both the spin and of the
orbital angular momentum $\mathbf{L}$ begins (left figure) about the
direction of the total angular momentum $\mathbf{J}$, which is due to the
spin-orbit interaction. Gravitational radiation carries away both energy and
angular momentum from the system, such that the direction of $\mathbf{J}$
stays unchanged. As a consequence the precessional orbit slowly shrinks and
the magnitude of $\mathbf{L}$ decreases. This is accompanied by a continuous
increase in the angle $\protect\alpha $ and a decrease in $\protect\beta $.
When the magnitudes of $\mathbf{L}$ and $\mathbf{S}$ become comparable
(middle figure), the precessional motions are much faster (for typical
values see Table \protect\ref{Table2}). In the typical mass ratio range $%
\protect\nu =1/3\div 1/30$ the magnitude of $\mathbf{L}$ becomes small as
compared to the magnitude of the spin, which is unchanged by gravitational
radiation (except the upper boundary of the range at $\protect\nu =1/3$ when 
$L$ and $S$\ are still comparable). Before reaching the innermost stable
orbit, the spin becomes almost aligned to the (original direction of the)
total angular momentum, and a new jet can form along this direction.
Therefore, for the typical mass ratio range the spin-flip phenomenon has
occurred in the inspiral phase and not much orbital angular momentum is left
to modify the direction of the spin during the plunge and ring-down. In the
regime in between the initial and final states the precessing jet acts as a
superwind, sweeping away the environment of the jets.}
\label{Fig1}
\end{figure*}

Dynamics becomes dissipative from 2.5 PN orders. Then gravitational
quadrupolar radiation carries away both energy and angular momentum. Orbital
eccentricity is dissipated faster, than the rate of orbital inspiral (Peters
1964), thus the orbit will circularize.\footnote{%
As we focus here on spin-flips, we will not dwell on possible recoil as a
result of the momentum loss due to gravitational radiation in the merger of
the two black holes (see for example Br{\"{u}}gmann 2008, Gonzalez et al.
2007a, b). The accuracy of determining the distance between two separate and
independent active black holes (Marcaide \& Shapiro 1983, Brunthaler et al.
2005) is reaching a precision, which may soon allow for recoil to be
measurable; no such evidence has been detected yet.} Considering circular
orbits and averaging over one orbit gives the following dissipative change
in $\mathbf{L}$:%
\begin{equation}
\mathbf{\dot{L}}^{gw}=-\frac{32G\mu ^{2}}{5r}\left( \frac{Gm}{c^{2}r}\right)
^{5/2}\mathbf{\hat{L}~,}  \label{Ldotdis}
\end{equation}%
where $\mathbf{\hat{L}}$ represents the direction of the orbital angular
momentum. Then the total change in $\mathbf{L}$ is given by the sum of Eqs. (%
\ref{Ldotconserv}) and (\ref{Ldotdis}).

The spin-induced quasi-precessional effects both modulate the dynamics and
they have an important effect on gravitational wave detection (see Lang \&
Hughes 2006, 2008, Racine 2008, Gergely \& Mik\'{o}czi 2008).

The dissipative dynamics, with the inclusion of the leading order SO
precessions and the dissipative effects due to gravitational radiation,
averaged over circular orbits, was discussed in detail in Apostolatos et al.
(1994) for the one-spin case $\mathbf{S_{2}}=0$ and equal masses $\nu =1$.
For the typical mass ratio $\nu \in \left( 1/30,~1/3\right) $, and keeping
only the leading order contributions in the $\nu $-expansion also gives $%
\mathbf{S_{2}}=0$ (the leading order contributions in $\mathbf{S_{2}}$ are
of order $\nu ^{2}$). In this subsection we will analyze in depth the
angular evolutions and the timescales involved.

As for any vector $\mathbf{X}$ with magnitude $X$ and direction $\mathbf{%
\hat{X}}$ one has $\mathbf{\dot{X}}=X\mathbf{\skew{0}{\dot}{\hat{X}}}+\dot{X}%
\mathbf{\hat{X}}$, the change in the direction can be expressed as $\mathbf{%
\skew{0}{\dot}{\hat{X}}}=\left( \mathbf{\dot{X}}-\dot{X}\mathbf{\hat{X}}%
\right) /X$. Also the identity $X^{2}=\mathbf{X}^{2}$ gives $\dot{X}=\mathbf{%
\hat{X}}\cdot \mathbf{\dot{X}}$. Then Eqs. (\ref{SdotJ})-(\ref{Ldotdis})
imply%
\begin{eqnarray}
\dot{S}_{1} &=&0~,  \nonumber \\
\mathbf{\skew{0}{\dot}{\hat{S}}_{1}} &=&{\frac{2G}{c^{2}r^{3}}}\mathbf{J}%
\times \mathbf{\hat{S}_{1}}\ ,  \nonumber \\
\dot{L} &=&-\frac{32G\mu ^{2}}{5r}\left( \frac{Gm}{c^{2}r}\right) ^{5/2}~, 
\nonumber \\
\mathbf{\skew{0}{\dot}{\hat{L}}} &=&{\frac{2G}{c^{2}r^{3}}}\mathbf{J\times 
\hat{L}}~.  \label{dynamics}
\end{eqnarray}%
The total angular momentum $\mathbf{J}$ is also changed by the emitted
gravitational radiation. As no other change occurs up the 2PN orders, $%
\mathbf{\dot{J}=}\dot{L}\mathbf{\hat{L}}$ and%
\begin{eqnarray}
\dot{J} &=&\dot{L}\left( \mathbf{\hat{L}}\cdot \mathbf{\hat{J}}\right) 
\mathbf{~,}  \nonumber \\
\mathbf{\skew{0}{\dot}{\hat{J}}} &=&\frac{\dot{L}}{J}\left[ \mathbf{\hat{L}}%
-\left( \mathbf{\hat{L}}\cdot \mathbf{\hat{J}}\right) \mathbf{\hat{J}}\right]
~.  \label{Jdot}
\end{eqnarray}%
Note that from the second Eq. (\ref{Jdot}) it is immediate that the
direction of $\mathbf{J}$ changes violently, whenever $J$ is small compared
to $\dot{L}$.

To leading order in $\nu $ , the vectors $\mathbf{L}$, $\mathbf{S}_{\mathbf{1%
}}$, and $\mathbf{J}$ form a parallelogram, characterized by the angles $%
\alpha =\cos ^{-1}\left( \mathbf{\hat{L}\cdot \hat{J}}\right) $ and $\beta
=\cos ^{-1}\left( \mathbf{\hat{S}}_{\mathbf{1}}\mathbf{\cdot \hat{J}}\right) 
$. From Eqs.(\ref{dynamics}) and (\ref{Jdot}), we obtain%
\begin{eqnarray}
\dot{\alpha} &=&-\frac{\dot{L}}{J}\sin \alpha >0~,  \label{alphadot} \\
\dot{\beta} &=&\frac{\dot{L}}{J}\sin \alpha <0~.  \label{betadot}
\end{eqnarray}%
In the latter equation, we have used that $\mathbf{\hat{S}}_{\mathbf{1}}%
\mathbf{\cdot \hat{L}=\cos }\left( \alpha +\beta \right) $.

Thus, we have found the following picture for the inspiral of the compact
binary after the transition radius. By disregarding gravitational radiation,
the SO precessions (\ref{Sdot}) and (\ref{Ldotconserv}) assure that the
vectors $\mathbf{L}$ and $\mathbf{S_{1}}$ are precessing about $\mathbf{J}$
(a fixed direction), but also about each other (then the respective axes of
precession evolve in time). Gravitational radiation slightly perturbs this
picture. The angle $\alpha +\beta $ between the orbital angular momentum and
the dominant spin stays constant during the inspiral, even with the
gravitational radiation taken into account. By contrast, the angle between $%
\mathbf{J}$ and$\ \mathbf{L}$ continuously increases, while the angle between%
$\ \mathbf{J}$ and $\mathbf{S_{1}}$ decreases with the same rate. This also
means that due to gravitational radiation, the vectors $\mathbf{L}$ and $%
\mathbf{S_{1}}$ do not precess about $\mathbf{J}$ any more in an exact
sense. They keep precessing about each other, however.

The change in the total angular momentum $\mathbf{\dot{J}=}\dot{L}\mathbf{%
\hat{L}}$ is about the orbital angular momentum, which in turn basically
(disregarding gravitational radiation) undergoes a precessional motion about 
$\mathbf{J.}$ This shows that the averaged change in $\mathbf{J}$ is along $%
\mathbf{J}$. This conclusion, however, depends strongly on whether the
precessional angular frequency $\Omega _{p}$ is much higher than the change
in the angles $\alpha $ and $\beta $. Indeed, if these are comparable, the
component perpendicular to $\mathbf{J}$ in the change $\mathbf{\dot{J}=}\dot{%
L}\mathbf{\hat{L}}$ will not average out during one precessional cycle, as
due to the increase of $\alpha $ it can significantly differ at the
beginning and at the end of the same precessional cycle, see Fig \ref{Fig1}.

The regime with $\Omega _{p}\gg \dot{\alpha}$ can be well approximated by a
precessional motion of both $\mathbf{L}$ and $\mathbf{S_{1}}$ about a fixed $%
\mathbf{\hat{J}}$, with the magnitudes of $\mathbf{L}$ and $\mathbf{J}$
slowly shrinking, the angle $\alpha $ slowly increasing and $\beta $ slowly
decreasing. As a result, during the inspiral, the orbital angular momentum
slowly turns away from $\mathbf{J}$, while $\mathbf{S_{1}}$ slowly
approaches the direction of $\mathbf{J}$. This regime is characteristic for
the majority of cases, and it was called simple precession in Apostolatos et
al. (1994).

Whenever $\Omega _{p}\approx \dot{\alpha}$, the conclusion of having $%
\mathbf{\dot{J}}$ in the direction of $\mathbf{\hat{J}}$ does not hold for
the average over one precession. This results in a change in the direction
of $\mathbf{J}$ in each precessional cycle. The evolution becomes much more
complicated (in fact no approximate analytical solution is known), and it
was called transitional precession in Apostolatos et al. (1994).

Let us see now when the two types of evolution typically occur. For this, we
note that the inspiral rate $\dot{L}/L$ is of the order 
\begin{equation}
\frac{\dot{L}}{L}\approx -\frac{32c^{3}}{5Gm}\varepsilon ^{4}\eta ~,
\end{equation}%
while the precessional angular velocity $\Omega _{p}=2GJ/c^{2}r^{3}$ gives
the estimate%
\begin{equation}
\Omega _{p}\mathbf{\approx }\frac{2c^{3}}{Gm}{\varepsilon }^{5/2}\eta \frac{{%
J}}{L}~.  \label{Omegapepeta}
\end{equation}%
Finally, the tilt velocity of $\mathbf{L}$ is of the order%
\begin{eqnarray}
\dot{\alpha} &\approx &\frac{32c^{3}}{5Gm}\varepsilon ^{7/2}\eta \nu \left( 
\frac{S_{1}}{J}\right) ^{2}\sin \left( \alpha +\beta \right)  \nonumber \\
&\approx &\frac{32c^{3}}{5Gm}\varepsilon ^{9/2}\eta \nu ^{-1}\left( \frac{L}{%
J}\right) ^{2}\sin \left( \alpha +\beta \right)  \label{alphadotepeta}
\end{eqnarray}
We have used 
\begin{equation}
\sin \alpha =\frac{S_{1}}{J}\sin \left( \alpha +\beta \right) \approx
\varepsilon ^{1/2}\nu ^{-1}\frac{L}{J}\sin \left( \alpha +\beta \right)
\label{albe}
\end{equation}
for the first and second expressions of $\dot{\alpha}$ , respectively,
together with Eq. (\ref{S1L}). For comparison, these are all represented in
Table \ref{Table2} for all three regimes characterized by $S_{1}/L\approx
0.3 $, $S_{1}\approx L$ and $S_{1}/L\approx 3$, respectively. 
\begin{table*}[tbp]
\centering%
\begin{minipage}{140mm}
\caption{Order of magnitude estimates for the inspiral rate $\dot{L}/L$,
angular precessional velocity $\Omega _{p}$ and tilt velocity $\dot{\protect%
\alpha}$ of the vectors $\mathbf{L}$ and $\mathbf{S}_{\mathbf{1}}$ with
respect to $\mathbf{J}$, represented for the three regimes with $L>S_{1}$, $%
L\approx S_{1}$ and $L<S_{1}$, characteristic in the domain of mass ratios $%
\protect\nu =0.3\div 0.03$. The numbers in brackets represent inverse time scales in seconds$^{-1}$, calculated for the typical
mass ratio $\protect\nu =10^{-1}$, post-Newtonian parameter $10^{-3}$, $%
10^{-2}$ and $10^{-1}$, respectively and $m=10^{8}M_{\odot }$ (then $%
c^{3}/Gm=2\times 10^{-3}$ s$^{-1}$).}
\begin{center}
\begin{tabular}{crlrlrl}
   		& 
$L>S_{1}$ 	&
		& 
$L\approx S_{1}$ & 
		&
$L< S_{1}$  & \\
\hline\hline
$-\dot{L}/L $ 					 & 
$\quad\frac{32c^{3}}{5Gm}\varepsilon^{4}\eta $  &
$\left(\approx 10^{-15}\right) $ 		 & 
$\quad\frac{32c^{3}}{5Gm}\varepsilon ^{4}\eta $ &
$\left(\approx 10^{-11}\right) $ 		 & 
$\quad\frac{32c^{3}}{5Gm}\varepsilon ^{4}\eta $ &
$\left(\approx 10^{-7}\right) $ \\
$\quad\Omega _{p}$ 				 & 
$\quad\frac{2c^{3}}{Gm}\varepsilon ^{5/2}\eta $ &
$\left(\approx 10^{-11}\right) $ 		 & 
$\quad\frac{2c^{3}}{Gm}\varepsilon ^{5/2}\eta \frac{J}{L}$ &
$\left(\approx 10^{-8}\frac{J}{L}\right) $ & 
$\quad\frac{2c^{3}}{Gm}\varepsilon ^{3}$		 &
$\left(\approx 10^{-5}\right) $ \\
$\quad\frac{\dot{\alpha}}{\sin(\alpha+\beta)} $ 				 & 
$\quad\frac{32c^{3}}{5Gm}\varepsilon ^{9/2}\frac{\eta}{\nu}$	 &
$\left(\approx 10^{-16}\right) $ 		 & 
$\quad\frac{32c^{3}}{5Gm}\varepsilon ^{9/2}\frac{\eta}{\nu}\frac{L^{2}}{J^{2}}$ &
$\left(\approx 10^{-11}\frac{L^{2}}{J^{2}}\right)$	     & 
$\quad\frac{32c^{3}}{5Gm}\varepsilon ^{7/2}\eta\nu$	     &
$\left(\approx 10^{-8}\right) $
\end{tabular}
\end{center}
\label{Table2}
\end{minipage}
\end{table*}

The numbers from the second line of Table \ref{Table2} demonstrate that for
the chosen typical example the precession timescale can get as short as a
day, going from 3000 years to three years to a day in the three columns
above. This last stage is obviously quite close to the plunge. From the
first line we can infer upper limits of how close the merger actually is, so
30 million years in Column 1, 300 years in Column 2, and a few months in
Column 3. As the inspiral rate increases in time, rather than being a
constant, these numbers are higher than the real values. The accuracy of the
third estimate is further obstructed by the fact that after ISO the plunge
follows, but as this comprises only a few orbits, the PN prediction can be
considered relevant as an order of magnitude estimate. By multiplying the
numbers in the third line with the precession timescales $\Omega _{p}^{-1}$
we actually get the relevant tilt angle, varying from 2 arcsec during one
precession ($6\times 10^{-4}$ arcseconds per year) at large separations to 3
arcmin per precession (per day) close to the ISO.

We see, that the rate of precession and the tilt velocity become comparable
in the $S_{1}\approx L$ epoch (in which $\varepsilon ^{1/2}\nu ^{-1}\approx
1 $) for 
\begin{equation}
\left[ \frac{16}{5}\sin \left( \alpha +\beta \right) \right] ^{-1/3}\frac{J}{%
L}\approx \left( \varepsilon ^{2}\nu ^{-1}\right) ^{1/3}\approx \varepsilon
^{1/2}\approx \nu ~,  \label{transitional}
\end{equation}%
that is, for the chosen numerical example $\nu =10^{-1}$ and for the square
bracket of order unity, this gives $J/L\approx 10^{-1}$ and the rate of $%
\dot{\alpha}\approx \Omega _{p}\approx 10^{-9}$ (this is still $100$ times
faster than the rate of inspiral).

The total angular momentum $J$ can be that small only if $\mathbf{L}$ and $%
\mathbf{S_{1}}$ are almost perfectly anti-aligned, thus $\alpha +\beta
\approx \pi -\delta $, $\delta \ll 1$, and~$L\approx S_{1}$. What does the
condition for transitional precession (\ref{transitional}) mean in terms of
their angle, how close should that be to the perfect anti-alignment? To
answer the question we note that%
\begin{equation}
\frac{J}{L}=\frac{L\cos \alpha +S_{1}\cos \beta }{L}\approx \allowbreak
\delta \sin \alpha ~,
\end{equation}%
Then, comparing with the estimates (\ref{albe}) and (\ref{transitional}) we
conclude, that transitional precession can occur only if the deviation from
the perfect anti-alignment is of the order of $\nu ^{3/2}$. This is a highly
untypical case in galactic black hole binaries.

\subsection{The limits of validity}

One might seriously question whether pushing the values of the parameters
beyond the range for which we use them would actually demonstrate that our
approximations cannot possibly be correct. We would like to address the
following two concerns specifically.

a) Is the high orbital angular momentum limit $L>>S_1$, obtained by a
sufficient increase of the separation of the two black holes a correct limit?

b) Is the extreme mass limit $\nu \rightarrow 0$, for which the treatment of
Section 5 may seem increasingly accurate, correct?

Concerning the limit (a): in any expansion with a small parameter one
condition always holds: the parameter must be small, and in any PN expansion
we can always reach a stage, for which the physics basis fails as
inadequate, as other additional physical processes become dominant, or for
which there is no observational support, despite the mathematical validity
of the expansion.

From among the effects affecting the directions of the spins the dynamics
discussed in Section 5 takes into account (1) the leading order conservative
effect, given by the precession due to spin-orbit coupling and (2) the
leading order dissipative effect due to gravitational radiation. Other
conservative and dissipative effects are neglected, being weaker. Meaningful
results can be traced from this model only when these assumptions hold. This
implies that the post-Newtonian parameter $\varepsilon $ varies between $%
10^{-3}$ and $10^{-1}$, corresponding to orbital separations of 500
Schwarzschild radii $=0.005$ pc to 5 Schwarzschild radii, given a $10^{8}$ $%
M_{\odot }$ black hole. This is emphasized in the paragraph following Eq. (%
\ref{rep}).

The "initial" and "final" phases in the dynamics described above therefore
refer to a well-defined range of orbital separation, they are not arbitrary.
The choice for the initial state is further justified by the discussion of
Section 4. One cannot apply the dynamics discussed above at arbitrarily
large distances, where the orbital momentum indeed would dominate, simply
because the dynamics is no longer valid there. At larger separations the
leading order dissipative effect is due to dynamical friction, thus the
discussion of the previous two subsections does not apply.

Concerning (b): according to the summary presented in Table 1, there are
three possibilities in the PN regime, where the dynamics discussed above
holds: (i) from mass ratios ${\nu }$ from $1$ to $1/3$ the orbital angular
momentum dominates throughout the whole range of separations between $0.005$
pc and 5 Schwarzschild radii, as noted; (ii) from mass ratios $\nu $ from $%
1/3$ to $1/30$ initially the orbital angular momentum dominates over the
spin, but their ratio is reversed at the final separation; (iii) from mass
ratios $\nu $ smaller than $1/30$ the spin is dominant throughout the
process.

Our claim is that the spin-flip is produced only if the total angular
momentum (whose direction stays unchanged) initially is dominated by the
orbital angular momentum, finally by the spin, thus only for the case (ii).

In the dynamics presented in the previous two subsections we have neglected
the second spin. As even for the highest mass ratio $\nu =1/3$ in the regime
(ii) the second spin is 1 order of magnitude smaller than the leading spin,
we consider this assumption justified. With decreasing mass ratios it
becomes increasingly accurate to neglect the second spin, as according to
Eq. (7) the ratio of the spins goes with $\nu ^{2}$.

However, not all results of the previous subsection become increasingly
accurate with a decreasing mass ratio. We emphasize, what is different in
case (iii) as compared to (ii). The difference is in the initial conditions,
which allow to obtain a spin-flip in case (ii), but not in (iii).
Mathematically, the difference between these two cases can be seen from Eq.
(33), showing that the angular tilt velocity of the dominant spin scales
with $\nu ^{2}$. For the extreme mass ratios $\nu <1/30$, e.g., Table 1, the
spin dominates over the orbital angular momentum throughout the whole PN
regime. Therefore, the ratio $S_{1}/J$ is of order unity. With decreasing $%
\nu $ however the change in the direction of the spin, represented by $\dot{%
\alpha}$ goes fast to zero, thus no spin-flip is produced in the PN regime
for extreme mass ratios.

At the end of this subsection we derive an analytical expression relating $%
\alpha $ to the conserved $\alpha +\beta $, the evolving PN parameter $%
\varepsilon $ and the mass ratio $\nu $. From the Figure \ref{Fig1} $J=L\cos
\alpha +S_{1}\cos \beta $. By introducing the angle $\alpha +\beta $ and
employing the estimate (\ref{S1L}) we obtain 
\begin{equation}
\frac{J}{L}\approx \left[ 1+\varepsilon ^{1/2}\nu ^{-1}\cos \left( \alpha
+\beta \right) \right] \cos \alpha +\varepsilon ^{1/2}\nu ^{-1}\sin \left(
\alpha +\beta \right) \sin \alpha ~.
\end{equation}%
Inserting this into the second expression (\ref{albe}) and rearranging we
find%
\begin{equation}
\frac{\sin 2\alpha }{1+\cos 2\alpha }\approx \frac{\sin \left( \alpha +\beta
\right) }{\varepsilon ^{-1/2}\nu +\cos \left( \alpha +\beta \right) }~.
\end{equation}%
For an initial configuration of $0.005$ pc (such that $\varepsilon \equiv
\varepsilon ^{\ast }=10^{-3}$) and mass ratio $\nu =10^{-1}$, the initial
misalignment between $\mathbf{L}$ and $\mathbf{J}$ is $\alpha
_{initial}\approx 18^{\circ },~10^{\circ },~0^{\circ }$ for the dominant
spin in the plane of orbit, spanning $45^{\circ }$ with the plane of orbit
and perpendicular to the plane of orbit (such that $\alpha +\beta =90^{\circ
},~45^{\circ },~0^{\circ }$), respectively. Then $\beta _{initial}=72^{\circ
},~35^{\circ },~0^{\circ }$. For the same mass ratio and relative
configurations, the angle $\alpha $ at the end of the PN epoch (at $%
\varepsilon =10^{-1}$) becomes $\alpha _{final}\approx 73^{\circ
},~35^{\circ },~0^{\circ }$,\ respectively. This can be translated into a
misalignment between $\mathbf{S}_{\mathbf{1}}$ and $\mathbf{J}$ of $\beta
_{final}=17^{\circ },~10^{\circ },~0^{\circ }$, and a spin-flip of $\Delta
\beta =55^{\circ },~25^{\circ },~0^{\circ }$, respectively.

\subsection{Summary}

In the typical range of mass ratios $\nu =0.03\div 0.3$ the initial
condition $L>S_{1}$ is always transformed into $S_{1}<L$, but the transition
is very rarely accompanied by the so-called transitional precession. In all
other cases the precession is simple. As the precession angle of the
dominant spin is decreasing in time from the given initial value to a small
value, the precessional cone becomes narrower in time. At the end of the
inspiral phase the dominant spin $\mathbf{S}_{\mathbf{1}}$ will point
roughly along $\mathbf{J.}$ This means that a spin-flip has occurred during
the post-Newtonian evolution, already in the inspiral phase of the merger.
On the other hand, as the inspiral phase ends with $L<S_{1}$, irrespective
of what happens in the next phases, during the plunge and ring-down, $L$ is
not high enough to cause additional significant spin-flip.

For smaller mass ratios (for extreme mass ratio mergers) the larger spin
already dominates the total angular momentum from the beginning of the
inspiral, thus no spin-flip will occur by the mechanism presented here.
Alternatively, from the second expression (\ref{alphadotepeta}) one can see
that the rate of tilt of the spin decreases with $\nu ^{2}$, thus it goes
fast to zero in the extreme mass ratio case. However, as we argued in
Section 2, such mass ratios are less typical for galactic central SMBH
mergers. This also shows that an infalling particle will not change the spin
of the supermassive black hole.

For the (again untypical) equal mass SMBH mergers the orbital angular
momentum stays dominant until the end of the inspiral phase. In this case,
however the possibility remains open to have a spin-flip later on, during
the plunge.

\section{Discussion}

The considerations from this paper lead to the following time sequence for
the transient feeding of a SMBH including a merger with another SMBH.

\textit{First:} Two galaxies with central black holes approach each other to
within a distance where dynamical friction keeps them bound, spiraling into
each other. If there is cool gas in either one, it can begin to form stars
rapidly, along tidal arms. The galactic central supermassive SMBH binary
influences gas dynamics and star formation activity also in the nuclear gas
disk, due to various resonances between gas motion and SMBH binary motion
(Matsui et al. 2006), creating some characteristic structures, such as
filament structures, formation of gaseous spiral arms, and small gas disks
around SMBHs. If either galaxy happens to have radio jets, then due to the
orbital motion, these jets get distorted and form the Z-shape (Gopal-Krishna
et al. 2003, Zier 2005).

\textit{Second: }The central regions in each galaxy begin to act as one
unit, in a sea of stars and dark matter of the other galaxy. During this
phase, as the cool gas from the other partner typically has low angular
momentum with respect to the receiving galaxy, the central region can be
stirred up, and produce a nuclear starburst (Toomre and Toomre 1972). The
central black hole can get started to be fed at a high rate, but its
emission will be submerged in all the far-infrared emission from the gas and
dust heated by the massive stars produced in the starburst. In this case,
there is dynamical friction, which can act so as to select certain
symmetries, such as corotation, counter rotation, or rotation at $90^{\circ
} $ (as in NGC2685, a polar ring galaxy; Richter et al. 1994).

\textit{Third: }The black holes begin to lose orbital angular momentum due
to the interaction with the nearby stars (Zier and Biermann 2001, 2002). \
Other mechanisms for angular momentum loss are also known (Sesana et al.
2006, 2007a,b, Alexander 2007, Hayasaki 2008). The two black holes approach
each other to that critical distance where the interaction with the stars
and the gravitational radiation remove equivalent fractions of the orbital
angular momentum. Then, as shown in this paper, the spin axes tumble and
precess. This phase can be identified with the apparent superdisk, as the
rapidly precessing jet produces the hydrodynamic equivalent of a powerful
wind, by entraining the ambient hot gas, pushing the two radio lobes apart
and so giving rise the a broad separation (Gopal-Krishna et al. 2003, 2007,
Gopal-Krishna \& Wiita 2006). Gopal-Krishna \& Wiita (2006) emphasize the
apparent asymmetry, which we propose to attribute to line-of-sight effects
and the distortion due to the recent merger. The base of the radio structure
is so broad and so asymmetric, that the central AGN will appear to be offset
from the projected center of gap. The recent arguments of Worrall et al.
(2007) seem to be consistent with this point of view. The spin direction of
the combined two black holes is preserved, although the strength of the spin
decreases. As during simple precession the total angular momentum shrinks
considerably, but its direction is conserved, on the other side the
magnitude of the spin stays constant, this means that the orbital angular
momentum shrinks. For comparable mass binaries it will be still higher than
the spin at ISO (therefore the dynamics below ISO, which can be analyzed
only numerically, should be responsible for any spin-flip in the comparable
masses case). For extreme mass ratio binaries the result of the shrinkage of
the orbital angular momentum is $L<S$ at ISO. Therefore, the spin at ISO
should be roughly aligned with the direction of $\mathbf{J}=\mathbf{L}+%
\mathbf{S}$, which (as initially $\mathbf{L}$ was dominant), is close to the
direction of the initial $\mathbf{L}$.

In certain cases, especially for equal masses of the two black holes, a
strong recoil has been found (Gonz{\'{a}{}}lez et al. 2007a, b). However, as
we noted earlier, the equal mass ratio is untypical.

\textit{Fourth:} The two black holes actually merge, and the merged black
hole keeps the spin axis from the orbital angular momentum of the previously
existing binary, whenever the mass ratio is relatively large. In the case
that the mass ratio is between $1:1$ and $1:3$, then even at the innermost
stable orbit a substantial fraction of the orbital angular momentum can
survive, possibly leading to a spin-flip later on. This very short phase
should be accompanied by extreme emission of low-frequency gravitational
waves. The final stage in this merger leads to a rapid increase in the
frequency of the waves, called \textquotedblleft chirping", but this
chirping will depend on the angles involved. The angle between the orbital
spin of the combined two black holes, and intrinsic spin of the more massive
black hole influences the highest frequency of the chirp; for a large angle
this frequency will be lower than for a small angle between the two spins.
Whether there is another observable feature, such as the induced decay of
heavy dark matter particles, from the merger of the two black holes at that
event such as speculated by Biermann \& Frampton (2006) is not clear at this
time.

\textit{Fifth:} Now the newly oriented more massive merged black hole starts
its accretion disk and jet anew, boring a new hole for the jets through its
environment. This stage can be identified perhaps with giga hertz peaked
radio sources (GPS). If the new jet points at the observer, then 3C147 may
be one example (Junor et al. 1999).

\textit{Sixth:} The newly oriented jets begin to show up over some kpc, and
this corresponds to the X-shaped radio galaxies, while the old jets are
fading but still visible. This also explains many of the compact steep
spectrum sources, with disjoint directions for the inner and outer jets.

\textit{Seventh: }The old jets have faded, and are at most visible in the
low radio frequency bubbly structures, such as seen for the Virgo cluster
region around M87 (Owen et al. 2000). The feeding is slowing down, and there
is no longer an observable accretion disk, but probably only an
advection-dominated disk. However, a powerful jet is still there, although
below or even far below the maximal power. The feeding is still from the
residual material stemming from the merger.

\textit{Eighth:} The feeding of the black hole is down to catching some gas
out of a common red giant star wind as presumably is happening in our
Galactic center. This stage seems to exist for all black holes, even at very
low levels of activity (e.g., Perez-Fournon \& Biermann 1984, Elvis et al.
1984, Nagar et al. 2000).

If this concept described here is true, then the superdisk radio galaxies
should have large outer distortions in their radio images, that may be
detectable at very high sensitivity, as they should correspond to recently
active Z-shaped sources. Also, the superdisk should be visible in X-rays,
although if the cooling is efficient the temperature may be relatively low.
Table 2 suggests that the merger is imminent, if the precession of the jet
is measurable within a few years, and the opening angle of the precession is
much narrower than the wind cone, reflecting the earlier longer time
precession (see Gopal-Krishna et al. 2007). Therefore, with very sensitive
radio interferometry it might be possible to detect the underlying jet
despite its rapid precession, although immediately before the actual merger
the feeding of the jet will be turned off.

As more and more pieces of evidence suggest that AGNs are the sources of
ultra-high energy cosmic rays (Biermann \& Strittmatter 1987, Biermann et
al. 2007) we need to ask what we could learn next. Clearly, after a
spin-flip, the new relativistic jet bores through a new environment, with
lots of gas, and so suffers a strong decelerating shock. In such a shock
particles are accelerated to maximal energies, and at the same time, as they
leave the shock region interact with all that interstellar gas. Therefore,
such sites are primary sources for any new particles, such as high energy
neutrinos (Becker et al. 2007). Such discoveries may well be possible long
before we detect the low frequency gravitational waves from the black hole
merger. As at such high energy neutrinos travel straight across the
universe, and suffer little loss other than from the adiabatic expansion of
the universe, the black holes resulting from a merger of two black holes,
with subsequent spin-flip, will be primary targets for searches for
ultra-high energy neutrinos, and perhaps other photos and particles at
extreme energies.

\section{Concluding Remarks}

Whereas it has been questioned in the past whether the central SMBHs of
merging galaxies will be able to actually merge or their approach will stall
(due to the process of loss-cone depletion) at a distance where dissipation
through gravitational radiation is not yet efficient (for a review of these
considerations see Merritt \& Miloslavljevi\'{c} 2005), the role of the
dynamical friction as bringing close the SMBHs to the transition radius,
from where gravitational radiation undertakes the control of the dissipative
process has been recently confirmed (Zier 2006) and also complementary
mechanisms were proposed (Alexander 2007, Sesana et al. 2006, 2007a,b,
Hayasaki 2008). The space mission LISA is predicted to detect the merger of
SMBHs. The statistical arguments of Rottmann (2001), using radio
observations, suggest that all strong central activity in galaxies may
involve a merger of two black holes. Therefore, we have assumed in this
paper that whenever galaxies merge, so will do their central SMBHs. Even if
there would be exceptions under this rule, this would reflect only in the
inclusion of an overall factor $\lesssim 1$ in the number of mergers of
SMBHs as compared to the number of mergers of galaxies, derived in Section
2, which would not affect the mass ratio estimates of our paper.

Guided by reasonable and simple assumptions we have shown that binary
systems of \textit{SMBH binaries formed by galaxy mergers typically have a
mass ratio range between }$1/3$\textit{\ and }$1/30$. Following this, we
have proven that for the typical mass ranges a combination of the SO
precession and gravitational radiation driven dissipation produces \textit{%
the spin-flip of the dominant black hole already in the inspiral phase},
except for the particular configuration of the spin perpendicular to the
orbital plane. During this process the magnitude of the spin is unchanged,
therefore \textit{the merger of a high spin} (and high rotation parameter) 
\textit{black hole with the smaller black hole results in a similar high
spin state} at the end of the inspiral phase. These are the main results of
our paper.

There is a related discussion undergoing in the literature, whether the high
spin of SMBHs is produced by prolonged accretion phases or by frequent
mergers. Even a scenario, where the SMBHs have typically low spin (King \&
Pringle 2006) was advanced, based on the assumption of short periods of
small accretion from random directions. Hughes \& Blandford (2003),
extrapolating results from $\nu =q^{-1}\ll 1$ binaries to comparable masses,
have shown that mergers spin-down black holes. Volonteri et al. (2005) have
studied the distribution of SMBH spins under the combined action of
accretion and mergers, and found that the dominant spin-up effect is by gas
accretion. Recently, Berti \& Volonteri (2008) have considered the problem
of mergers by taking into account improvements in the numerical general
relativistic methods (Pretorius 2007), and a recent semianalytical formula,
which gives the final spin in terms of the initial dimensionless spins, mass
ratio, and relative angles of orbital angular momentum and spins (Rezzolla
et al. 2008a,b,c, Barausse \& Rezzolla 2009). They have found that mergers
can result in a high spin end state only if the dominant spin is aligned
with the orbital angular momentum of the system (thus the smaller mass
orbits in the equatorial plane of the larger). Their considerations extend
from comparable masses to mass ratios of $1/10$. However, Berti and
Volonteri (2008) neglected the angular momentum exchange and transport
between black hole, jet, and inner accretion disk by magnetic fields (see,
e.g., Blandford 1976); this may modify or even sharpen the conclusions.

We can add three remarks to this discussion. First, we have shown by
analytical means, that for the typical mass ratio range the inspiral phase
ends with a considerably lower value of the orbital angular momentum
compared to the spin (see the last picture in Figure \ref{Fig1}). A
heuristic argument then shows that such a small angular momentum could not
significantly change the \textit{direction} of the spin during the next
phases of the merger. Apart from this small orbital angular momentum, the
problem being axially symmetric, \textit{we do not expect significant
further spin-flip due to gravitational radiation in the last stages of the
inspiral}.

Second, the configuration of \textit{orbital angular momentum aligned with
the dominant spin is not a preferred one} in the gravitational radiation
dominated post-Newtonian regime. It is not clear yet whether such an
alignment could be the by-product of previous phases of the inspiral, when
dynamical friction (Zier \& Biermann 2001), three-body interactions (Sesana
et al. 2006, 2007a,b), relaxation processes due to cloud-star interactions
(Alexander 2007), three disk model accretion (Hayasaki 2008), and other
possible mechanisms occur. Since the stellar system is often slightly
flattened, differential dynamical friction could produce the near alignment
necessary to allow very high spin after a merger.

Third, the \textit{magnitude} of the spin is practically unchanged in the
inspiral phase, discussed here. This is because the loss in the spin vector
by gravitational radiation, a second PN order effect, calculated from the
Burke-Thorne potential (Burke 1971), is perpendicular to the spins, yielding
another precessional effect (Gergely et al. 1998c). Below ISO this estimate
should break down, as indicated by numerical simulations reporting on
various fractions of the spin radiated away. In this context we want to
emphasize the \textit{unchanged magnitude of the spin during the inspiral},
as important initial data for the numerical evolution during the plunge and
ring-down.

We also mention here the results of the numerical relativity community
showing a considerable recoil of the merged SMBH in particular cases, mostly
for equal masses and peculiar configurations of the angular momenta (Br{\"{u}%
}gmann 2008, Gonzalez et al. 2007a, b, Koppitz et al. 2007). It has also
been shown that the recoil regulates the SMBH mass growth, as the SMBH
wanders through the host galaxy for $10^{6}\div 10^{8}$ years (Blecha and
Loeb 2008). According to the empirical formula of Campanelli et al. (2007a,
see also Lousto \& Zlochower 2009) the recoil velocity scales with $%
q^{-2}/\left( 1+q^{-1}\right) ^{2}\left( 1+q^{-1}\right) $, which for $%
q^{-1}=\nu \ll 1$ reduces to a scaling with $q^{-2}$. Therefore, \textit{we
do not expect significant recoil effects in the typical mass ratio range of
the SMBH mergers}.

We suggest that the precessional phase of the merger of two black holes,
occurring prior to the spin-flip, is visible as a superdisk in radio
galaxies (Gopal-Krishna et al. 2007). The \textit{precessing jet appears as
a superwind separating the two radio lobes} in the final stages of the
merger. According to our model such radio galaxies are candidates for
subsequent SMBH mergers. Further observations and theoretical work may be
capable of identifying such candidates likely to merge, and determine the
timescale for this to happen. The restart of powering a relativistic jet
(after the spin-flip and merging) will produce ultra-high energy hadrons,
neutrinos and other particles.

Based on the estimates given in Table 2 for the precessional and inspiral
timescales, we can say the following. If we were to observe a precession
timescale of three years in a superdisk radio galaxy, we could confidently
predict a plunge in about 300 years, which should be observable. Faster
precession timescales would take some effort to identify. However, if we
were able to even identify a precession timescale of days to weeks, then the
plunge would be predicted to happen a few months to a few years thence:
powerful gravitational waves at very low frequency would then be emitted.

The picture developed here differs from that in Wilson \& Colbert (1995) in
that we do not identify just the rare mergers of two massive black holes of
about equal masses with radio galaxies and radio quasars. We intend to
revisit the interactions with the stars (Zier et al. 2009, in preparation),
discuss the spin of the black holes in another work (Kov\'{a}cs et al. 2009,
in preparation) developed from Du\c{t}an \& Biermann (2005), finally to work
out quantitatively the relation of the merger of black holes and the
statistics of radio galaxies (Gopal-Krishna et al. 2009, in preparation).

\section{Acknowledgements}

We are grateful for discussions with Gopal-Krishna and C. Zier. P.L.B.
acknowledge further discussions with J. Barnes, B. Br{\"{u}}gmann, and G. Sch%
{\"{a}}fer. L.\'{A}.G. was successively supported by OTKA grants 46939,
69036, the J\'{a}nos Bolyai Grant of the Hungarian Academy of Sciences, the
London South Bank University Research Opportunities Fund and the Pol\'{a}nyi
Program of the Hungarian National Office for Research and Technology (NKTH).
Support for P.L.B. was from the AUGER membership and theory grant 05 CU 5PD
1/2 via DESY/BMBF and VIHKOS. The collaboration between the University of
Szeged and the University of Bonn was via an EU Sokrates/Erasmus contract.


\begin{thebibliography}{999}
\bibitem{1240} Alexander, T., in \textit{2007 STScI Spring Symp.: Black
Holes"}, eds, M. Livio \& A.M. Koekemoer, (Cambridge, Cambridge University
Press), in press (arXiv:0708.0688)

\bibitem{1244} Antonucci, R.R.J., Miller, J.S., \textit{Astrophys. J.} 
\textbf{297}, 621 - 632 (1985) 

\bibitem{1253} Apostolatos T.A., \textit{Phys. Rev. D }\textbf{52}, 605
(1995) 

\bibitem{1256} Apostolatos T.A., \textit{Phys. Rev. D }\textbf{54}, 2438
(1996) 

\bibitem{1249} Apostolatos T.A., Cutler C., Sussman G.J., Thorne K.S., 
\textit{Phys. Rev. D} \textbf{49}, 6274 (1994) 

\bibitem{} Barausse, E. Rezzolla, L., arXiv:0904.2577V1 [gr-qc] (2009) 

\bibitem{1259} Barker\ B.M., O'Connell R.F., \textit{Phys. Rev. D }\textbf{12%
}, 329 (1975)

\bibitem{1262} Barker B.M., O'Connell R.F., \textit{Gen. Relativ. Gravit.} 
\textbf{2}, 1428 (1979)

\bibitem{1269} 
Barnes, J.E., in \textit{Proc. of the 4th Sci. Meet. of the Span. Astron.
Soc. (SEA)} 2000, Highlights of Span. Astrophys. II. ed. J. Zamorano, J.
Gorgas, \& J. Gallego (Dordrecht: Kluwer), 85 

\bibitem{1265} Barnes, J.E., Hernquist, L., \textit{Annual Rev. of Astron.
\& Astrophys.} \textbf{30}, 705 (1992) 

\bibitem{1275} Barth A.J., Greene J.E., Ho L.C., \textit{Astrophys. J.
Letters} \textbf{619}, L151 (2005) 

\bibitem{1280} Becker J.K., Gro\ss\ A., M\"{u}nich K., Dreyer J., Rhode W.,
Biermann P.L., \textit{Astropart. Phys.} \textbf{28}, 98 (2007) 

\bibitem{1285} Benson A.J., D\v{z}anovi\'{c} D., Frenk C.S., Sharples R., 
\textit{Mon. Not. Roy. Astron. Soc. }\textbf{379}, 841-866 (2007) 

\bibitem{1289} 
Berczik P., Meritt D., Spurzem R., \textit{Astrophys. J. Letters} \textbf{633%
}, 680 - 687 (2005) 

\bibitem{1297} 
Berczik P., et al., \textit{Astrophys. J. Letters} \textbf{642}, L21 - L24
(2006)

\bibitem{1304} Berti E., Volonteri M., \textit{Astrophys. J.} \textbf{684},
822 (2008) 

\bibitem{1307} Biermann P.L., Strittmatter P.A., \textit{Astrophys. J.} 
\textbf{322}, 643 (1987) 

\bibitem{1311} Biermann P.L., Strom R.G., Falcke H., \textit{Astron. \&
Astroph.} \textbf{302}, 429 (1995) 

\bibitem{1317} Biermann P.L., Chirvasa M., Falcke H., Markoff S., Zier Ch.,
invited review at the Paris Conference on Cosmology, June 2000, in
Proceedings, Eds. N. Sanchez, H. de Vega, p. 148 - 164 (2005);
astro-ph/0211503 

\bibitem{1323} Biermann P.L., Frampton P.H., \textit{Physics Letters B} 
\textbf{634}, 125 - 129 (2006) 

\bibitem{} Biermann P.L., Hall D.S., \textit{Astron. \& Astroph.} \textbf{27}%
, 249 - 253 (1973). 

\bibitem{1328} Biermann P. L., Isar P.G., Mari\c{s} I.C., Munyaneza F., Ta%
\c{s}c\u{a}u O., "Origin and physics of the highest energy cosmic rays: What
can we learn from Radio Astronomy ?\textit{"}, invited lecture at the Erice
meeting June 2006, editors M.M. Shapiro, T. Stanev, J.P. Wefel, World
Scientific, p. 111 (2007); astro-ph/0702161

\bibitem{} Biermann P. L., Becker J. K., Caramete A., Curu\c{t}iu L., Engel
R., Falcke H., Gergely L. \'{A}., Isar P. G., Mari\c{s} I. C., Meli A.,
Kampert K.-H., Stanev T., Ta\c{s}c\u{a}u O., Zier C., "Active Galactic
Nuclei: Sources for ultra high energy cosmic rays?", invited review for the
Proceedings of the CRIS 2008 - Cosmic Ray International Seminar: Origin,
Mass, Composition and Acceleration Mechanisms of UHECRs, Malfa, Italy, Ed. A
Insolia, Elsevier 2009; arXiv: 0811.1848v3 [astro-ph]

\bibitem{1334} Binney J., Tremaine S., \textit{Galactic Dynamics}, Princeton
University Press (1987)

\bibitem{1337} Blanchet L., Buonanno A., Faye G., \textit{Phys. Rev.} 
\textbf{D} \textbf{74}, 104034 (2006); Erratum-ibid. \textbf{75}, 049903
(2007)

\bibitem{} Blandford R.D. \textit{Month. Not. Roy. Astr. Soc.} \textbf{176},
465 (1976)

\bibitem{1340} Blecha L., Loeb A., \textit{Month. Not. Royal Astron. Soc.} 
\textbf{390}, 1311 (2008) 

\bibitem{1343} Bouwens R.J., Illingworth G.D., \textit{Nature} \textbf{443},
189 - 192 (2006) 

\bibitem{1357} 
Br{\"{u}}gmann B., Gonzalez J., Hannam M., Husa S., Sperhake U., \textit{%
Phys. Rev. }D \textbf{77}, 124047 (2008) 

\bibitem{1367} Brunthaler A., Reid M.J., Falcke H., Greenhill L.J., Henkel
C., \textit{Science} \textbf{307}, 1440 - 1443 (2005) 

\bibitem{1373} Burke W.L., \textit{J. Math. Phys.} \textbf{12}, 401 (1971)

\bibitem{1375} 
Campanelli M., Lousto C.O., Zlochower Y., Merritt D. \textit{Astrophys. J.} 
\textbf{659}, L5 (2007a) 

\bibitem{1380} 
Campanelli M., Lousto C.O., Zlochower Y., Krishnan B., Merritt D., \textit{%
Phys. Rev.} \textbf{D} \textbf{75}, 0640030 (2007b) 

\bibitem{1386} Chini R., Kreysa E., Biermann P.L., \textit{Astron. \&
Astroph.} \textbf{219}, 87-97 (1989a) 

\bibitem{1389} Chini R., Biermann P.L., Kreysa E., Gem{\"u}nd H.-P., \textit{%
Astron. \& Astroph. Letters} \textbf{221}, L3 - L6 (1989b). 

\bibitem{1393} Chirvasa M., Diploma thesis: "Gravitational Waves during the
mergers of rotating black holes", Bonn Univ. (2001) 

\bibitem{1397} Donea A.C., Biermann P.L., \textit{Astron. \& Astroph.} 
\textbf{316}, 43 (1996) 

\bibitem{1403} Du\c{t}an I., Biermann P.L., in the proceedings of the
International School of Cosmic Ray Astrophysics (14th course): "Neutrinos
and Explosive Events in the Universe", Ed. T. Stanev, published by Springer,
Dordrecht, The Netherlands, p.175 (2005), astro-ph/0410194 

\bibitem{1409} Elvis M., Soltan A., Keel W.C., \textit{Astrophys. J.} 
\textbf{283}, 479 - 485 (1984) 

\bibitem{1413} Faber S.M., Tremaine S., Ajhar E.A., et al. \textit{Astron. J.%
} \textbf{114}, 1771 (1997) 

\bibitem{1422} Falcke H., Biermann P.L., \textit{Astron. \& Astroph.} 
\textbf{293}, 665 (1995a) 

\bibitem{1428} Falcke H., Biermann P.L., \textit{Astron. \& Astroph.} 
\textbf{308}, 321 (1995b) 

\bibitem{1432} Falcke H., Biermann P.L., \textit{Astron. \& Astroph.} 
\textbf{342}, 49 - 56 (1999) 

\bibitem{1439} Falcke H., Malkan M.A., Biermann P.L., \textit{Astron. \&
Astroph.} \textbf{298}, 375 (1995) 

\bibitem{1445} Falcke H., Sherwood W., Patnaik A.R., \textit{Astrophys. J.} 
\textbf{471}, 106 (1996) 

\bibitem{1451} Faye G., Blanchet L., Buonanno A., \textit{Phys. Rev.} 
\textbf{D} \textbf{74}, 104033 (2006).

\bibitem{1454} 
Ferrarese L., Cote P., Blakeslee J.P., Mei S., Merritt D., West M.J., in 
\textit{IAU Sympos.}, \textbf{238}, in press (2006a); astro-ph/0612139 

\bibitem{1461} Ferrarese L. et al., \textit{Astrophys. J.} \ \textit{Suppl.}%
\ \textbf{164}, 334 (2006b)

\bibitem{1464} Flanagan E.E., Hinderer T., \textit{Phys. Rev. D} \textbf{75}%
, 124007 (2007) 

\bibitem{1467} Gergely L.\'{A}., Perj\'{e}s Z.. Vas\'{u}th M., \textit{Phys.
Rev.} \textbf{D} \textbf{57}, 876 (1998a)

\bibitem{1470} Gergely L.\'{A}., Perj\'{e}s Z.. Vas\'{u}th M., \textit{Phys.
Rev.} \textbf{D} \textbf{57}, 3423 (1998b)

\bibitem{1473} Gergely L.\'{A}., Perj\'{e}s Z.. Vas\'{u}th M., \textit{Phys.
Rev.} \textbf{D} \textbf{58}, 124001 (1998)

\bibitem{1476} Gergely L.\'{A}., \textit{Phys. Rev.} \textbf{D} \textbf{61},
024035 (2000a)

\bibitem{1479} Gergely L.\'{A}., \textit{Phys. Rev.} \textbf{D} \textbf{62},
024007 (2000b)

\bibitem{1482} Gergely L.\'{A}., Keresztes Z., \textit{Phys. Rev.} \textbf{D}
\textbf{67}, 024020 (2003)

\bibitem{} Gergely L.\'{A}., Mik\'{o}czi B., \textit{Phys. Rev. }\textbf{D} 
\textbf{79}, 064023 (2009) 

\bibitem{1485} Ghez A. M., Salim S., Hornstein S.D., et al., \textit{%
Astrophys. J.} \textbf{620}, 744 - 757 (2005) 

\bibitem{1494} Gilmore G., Wilkinson M., Kleyna J., Koch A., Wyn Evans N.,
Wyse R.F.G., Grebel E.K., presented at UCLA Dark Matter 2006 Conference,
March 2006, \textit{Nucl. Phys. Proc. Suppl.} \textbf{173}, 15 (2007) 

\bibitem{1502} Gonz{\'{a}}lez J.A. et al. \textit{Phys. Rev. Letters} 
\textbf{98}, 091101 (2007a) 

\bibitem{1514} Gonz{\'{a}}lez J.A., Hannam M.D., Sperhake U., Br{\"{u}}gmann
B., Husa S., \textit{Phys. Rev. Lett.} \textbf{98}, 231101 (2007b) 

\bibitem{1520} Gopal-Krishna, Wiita P.J., \textit{Astrophys. J.} \textbf{529}%
, 189 - 200 (2000) 

\bibitem{1523} Gopal-Krishna, Biermann P.L., Wiita P.J., \textit{Astrophys.
J. Letters} \textbf{594}, L103 - L106 (2003) 

\bibitem{1529} Gopal-Krishna, Biermann P.L., Wiita P.J., \textit{Astrophys.
J. Letters} \textbf{603}, L9 - L12 (2004) 

\bibitem{1535} Gopal-Krishna, Wiita P.J., Joshi S., \textit{Month. Not. Roy.
Astr. Soc.} \textbf{380}, 703 (2007) 

\bibitem{1539} Gopal-Krishna, Wiita P.J., invited talk at the 4th Korean
Workshop on high energy astrophysics (April 2006),
http://sirius.cnu.ac.kr/kaw4/presentations.htm 

\bibitem{1543} Gopal-Krishna, Zier Ch., Gergely L.\'{A}, Biermann P.L.,
(2009), in preparation

\bibitem{1546} Gott III J.R., Turner E.L., \textit{Astrophys. J.} \textbf{216%
}, 357 (1977)

\bibitem{1549} H\"{a}ring N., Rix H., \textit{Astrophys. J. Letters} , 
\textbf{604}, L89 (2004)

\bibitem{1552} Hayasaki K., to appear in \textit{Publications of the
Astronomical Society of Japan}, arXiv:0805.3408 (2008) 

\bibitem{1556} Hickson P., \textit{Astrophys. J.} , \textbf{255}, 382 (1982)

\bibitem{1558} Hughes S.A., Blandford R.D., \textit{Astrophys. J.} \textbf{%
585}, L101 (2003)

\bibitem{1561} Ioka K., Taniguchi K., \textit{Astrophys. J.} \textbf{537},
327 - 333 (2000) 

\bibitem{1566} Iye M. et al. \textit{Nature} \textbf{443}, 186 - 188 (2006) 

\bibitem{1577} Junor W., Salter C.J., Saikia D.J., Mantovani F., Peck A.B., 
\textit{Month. Not. Roy. Astr. Soc.} \textbf{308}, 955 - 960 (1999) 

\bibitem{1584} Kidder L., Will C., Wiseman A., \textit{Phys. Rev.} \textbf{D}
\textbf{47}, R4183 (1993)

\bibitem{1587} Kidder L., \textit{Phys. Rev.} \textbf{D} \textbf{52}, 821
(1995)

\bibitem{1589} King A.R., Pringle J.E., \textit{Month. Not. Roy. Astr. Soc.} 
\textbf{373}, L90 (2006)

\bibitem{1592} Klypin A., Zhao H.-S., Somerville R.S., \textit{Astrophys. J.}
\textbf{573}, 597 - 613 (2002) 

\bibitem{} Koppitz, M., Pollney, D., Reisswig, C., Rezzolla, L., Thornburg,
J., Diener, P., Schnetter E., \textit{Phys. Rev. Lett.} \textbf{99}, 041102
(2007) 

\bibitem{1597} Kormendy J., Richstone D., \textit{Annual Rev. of Astron. \&
Astrophys.} \textbf{33}, 581 (1995) 

\bibitem{} Kov\'{a}cs Z., Biermann, P.L., Gergely, \'{A}.L., "The maximal
spin of a black hole, disk and jet symbiotic system" in preparation (2009)

\bibitem{} Lang R.N., Hughes S.A., \textit{Phys. Rev.} \textbf{D} \textbf{74}%
, 122001 (2006). Errata, ibid. \textbf{D} \textbf{75}, 089902(E) (2007)

\bibitem{} Lang R.N., Hughes S.A., \textit{Astrophys. J.} \textbf{677}, 1184
(2008)

\bibitem{1602} Lauer T.R. et al., \textit{Astrophys. J.} \textbf{662}, 808
(2007)

\bibitem{1605} Lawrence A., Elvis M. \textit{Astrophys. J.} \textbf{256},
410 - 426 (1982) 

\bibitem{1609} Lousto C.O., Zlochower Y., \textit{Phys. Rev.} \textbf{D 79},
064018 (2009) 

\bibitem{1613} Lynden-Bell D., \textit{Month. Not. Roy. Astr. Soc.} \textbf{%
136}, 101 (1967) 

\bibitem{1618} Mahadevan R., \textit{Nature} \textbf{394}, 651 - 653 (1998) 

\bibitem{1622} 
Makino J., Funato Y., \textit{Astrophys. J.} \textbf{602}, 93 - 102 (2004) 

\bibitem{1627} Marcaide J.M., Shapiro I.I., \textit{Astron. J.} \textbf{88},
1133 - 1137 (1983) 

\bibitem{1632} Marecki A., Barthel P. D., Polatidis A., Owsianik I., \textit{%
Publ. Astron. Soc. Australia} \textbf{20}, 16 - 18 (2003) 

\bibitem{1637} 
Matsubashi T., Makino J., Ebisuzaki T.\textit{Astrophys. J.} \textbf{656},
879 - 896 (2007) 

\bibitem{1644} Matsui H., Habe A., Saitoh T.R., \textit{Astrophys. J.} 
\textbf{651}, 767 - 774 (2006) 

\bibitem{1652} 
Merritt D., in Proc. \textit{Coevolution of black holes and galaxies},
Cambridge U. Press, Ed. L.C. Ho (in press) (2003), astro-ph/0301257 

\bibitem{1657} 
Merritt D. \textit{Astrophys. J. Letters} \textbf{621}, L101 - L104 (2005) 

\bibitem{1662} 
Merritt D. \& Ekers R., \textit{Science} \textbf{297}, 1310-1313 (2002) 

\bibitem{1667} 
Merritt D., Mikkola S., Szell A., arXiv/0705.2745 

\bibitem{1676} Merritt D., Miloslavljevi\'{c} M., \textit{Living Rev.
Relativity} \textbf{8}, 8 (2005)

\bibitem{1679} Mik\'{o}czi B, Vas\'{u}th M, Gergely L.\'{A}., \textit{Phys.
Rev.} \textbf{D} \textbf{71}, 124043 (2005)

\bibitem{1682} 
Milosavljevi{\'{c}} M., Merritt D., "The Final Parsec Problem " , \textit{%
AIP Proc.} (in press), (2003a); astro-ph/0212270 

\bibitem{1687} 
Milosavljevi{\'{c}} M., Merritt D., \textit{Astrophys. J.} \textbf{596}, 860
- 878 (2003b) 

\bibitem{1694} Munyaneza F., Biermann P.L., \textit{Astron. \& Astroph.} 
\textbf{436}, 805 - 815 (2005) 

\bibitem{1698} Munyaneza F., Biermann P.L., \textit{Astron. \& Astroph.
Letters} \textbf{458}, L9 - L12 (2006) 

\bibitem{1702} Mushotzky, R. \textit{Astrophys. J.} \textbf{256}, 92 - 102
(1982) 

\bibitem{1708} Nagar N.M., Falcke H., Wilson A.S., Ho L.C., \textit{%
Astrophys. J.} \textbf{542}, 186 - 196 (2000) 

\bibitem{1713} O'Connell R.F., \textit{Phys. Rev. Letters} \textbf{93},
081103 (2004) 

\bibitem{1716} Owen F.N., Eilek J.A., Kassim N.E., \textit{Astrophys. J.} 
\textbf{543}, 611 (2000) 

\bibitem{1720} Perez-Fournon I., Biermann P.L., \textit{Astron. \& Astroph.
Letters} \textbf{130}, L13 - L15 (1984) 

\bibitem{1724} Peters P.C., \textit{Phys. Rev.} \textbf{136}, B1224 (1964)

\bibitem{1726} Peters P.C., Mathews S., \textit{Phys. Rev.} \textbf{131},
435\ (1963)

\bibitem{1729} Poisson E., \textit{Phys. Rev.} \textbf{D} \textbf{57}, 5287
(1998)

\bibitem{1732} Press W.H., Schechter P., \textit{Astrophys. J.} , \textbf{187%
}, 425 (1974)

\bibitem{1735} Pretorius F., in \textit{Relativistic Objects in Compact
Binaries: From Birth to Coalescence}, ed. Colpi et al., Springer Verlag,
Canopus Publishing Limited, arXiv:0710.1338 [gr-qc] (2007) 
\ 

\bibitem{1737} Racine E., \textit{Phys. Rev.} \textbf{D 78}, 044021 (2008)

\bibitem{} Rezzolla L., Barausse E., Dorband E. N., Pollney D., Reisswig
Ch., Seiler J. , Husa S., \textit{Phys. Rev.} \textbf{D 78}, 044002 (2008a) 

\bibitem{} Rezzolla L., Diener P., Dorband E.N., Pollney D., Reisswig Ch.,
Schnetter E., Seiler J., \textit{Astrophys. J.} \textbf{674}, L29 (2008b) 

\bibitem{} Rezzolla L, Dorband E.N., Reisswig Ch., Diener P., Pollney D.,
Schnetter E., Szilagyi B., \textit{Astrophys. J.} \textbf{679}, 1422 (2008c) 

\bibitem{1741} Richter O.-G., Sackett P.D., Sparke L.S., \textit{Astron. J.} 
\textbf{107}, 99 - 117 (1994) 

\bibitem{1746} Rieth R., Sch\"afer G., \textit{Class. Quantum Grav.} \textbf{%
14}, 2357 (1997)

\bibitem{1749} Roman S.-A., Biermann P.L., \textit{Roman. Astron J. Suppl.}, 
\textbf{16}, 147 (2006)

\bibitem{1752} Rottmann H., PhD thesis: "Jet-Reorientation in X-shaped Radio
Galaxies", Bonn Univ., 2001: (http://hss.ulb.uni-bonn.de/diss$\_$online/math 
$\_$nat$\_$fak/2001/rottmann$\_$helge/index.htm) 

\bibitem{1757} Ryan F., \textit{Phys. Rev.} \textbf{D 53}, 3064 (1996)

\bibitem{1759} Sanders D.B., Mirabel I.F., \textit{Annual Rev. of Astron. \&
Astrophys.} \textbf{34}, 749 (1996) 

\bibitem{1762} Sch{\"{a}}fer, G., \textit{Current Trends in Relativistic
Astrophysics}, Edited by L. Fern{\'{a}}ndez-Jambrina, L.M. Gonz{\'{a}}%
lez-Romero, Lecture Notes in Physics, vol. \textbf{617}, p. 195 (2005) 

\bibitem{1774} Sch{\"{o}}del R., Eckart A., \textit{Mem. Soc. Astron. Ital.} 
\textbf{76}, 65 (2005) 

\bibitem{1779} Sesana A., Haardt F., Madau P., \textit{Astrophys. J.} 
\textbf{651}, 392S (2006) 

\bibitem{1783} Sesana A., Haardt F., Madau P., \textit{Astrophys. J.} 
\textbf{660}, 546S (2007a) 

\bibitem{1787} Sesana A., Haardt F., Madau P., to appear in \textit{%
Astrophys. J.}; arXiv:0710.4301 (2007b) 

\bibitem{1791} Silk J., Takahashi T., \textit{Astrophys. J.} \textbf{229},
242 - 256 (1979) 

\bibitem{1795} Thorne, K. S., \textit{Proc. Royal Soc. London} \textbf{A 368}%
, 9 (1979) 

\bibitem{1798} Toomre A., Toomre J., \textit{Astrophys. J.} \textbf{178},
623 - 666 (1972) 

\bibitem{1801} 
Valtonen M.J., \textit{Month. Not. Roy. Astr. Soc.} 278, 186 (1996) 

\bibitem{1809} Vas\'{u}th M, Keresztes Z, Mih\'{a}ly A., Gergely L .\'{A}., 
\textit{Phys. Rev.} \textbf{D} \textbf{68}, 124006 (2003)

\bibitem{1812} Volonteri M., Madau P., Quataert E., Rees M. J., \textit{%
Astrophys. J.} \textbf{620}, 69 (2005)

\bibitem{1815} Wilson A.S., Colbert E.J.M., \textit{Astrophys. J.} \textbf{%
438}, 62 - 71 (1995)

\bibitem{1818} Worrall D.M., Birkinshaw M., Kraft R.P., Hardcastle M.J., 
\textit{Astrophys. J. Letters}, in press (2007); astro-ph/0702411 

\bibitem{1823} 
Yu Q., \textit{Class. Quantum Grav.} \textbf{20}, S55-S63 (2003) 

\bibitem{1828} Zier Ch., Biermann P.L., \textit{Astron. \& Astroph.} \textbf{%
377}, 23 - 43 (2001) 

\bibitem{1833} Zier Ch., Biermann P.L., \textit{Astron. \& Astroph.} \textbf{%
396}, 91 (2002) 

\bibitem{1838} Zier Ch., \textit{Month. Not. Roy. Astr. Soc.} \textbf{364},
583 (2005) 

\bibitem{1842} Zier Ch., \textit{Month. Not. Roy. Astr. Soc.} \textit{Lett.} 
\textbf{371}, L36 - L40 (2006) 

\bibitem{1852} Zier Ch., \textit{Month. Not. Roy. Astr. Soc.} \textbf{378},
1309-1327 (2007) 

\bibitem{1855} Zier Ch., Gergely L.\'{A}., Biermann P.L., (2009), in
preparation
\end{thebibliography}
\end{document}